\documentclass[prb,twocolumn,showpacs]{revtex4}
\usepackage{times}

\def\bK{\mathchoice{\mbox{\boldmath $\displaystyle K$}}
                   {\mbox{\boldmath $\textstyle K$}}
                   {\mbox{\boldmath $\scriptstyle K$}}
                   {\mbox{\boldmath $\scriptscriptstyle K$}}}

\def\bE{\mathchoice{\mbox{\boldmath $\displaystyle E$}}
           {\mbox{\boldmath $\textstyle E$}}
           {\mbox{\boldmath $\scriptstyle E$}}
           {\mbox{\boldmath $\scriptscriptstyle E$}}}

\def\imu{{\rm i}}
\def\euz{{\rm e}}
\def\eun{{\rm e}}
\def\RE{\mathop{\rm Re}\nolimits}
\def\IM{\mathop{\rm Im}\nolimits}
\def\sn{\mathop{\rm sn}\nolimits}
\def\cn{\mathop{\rm cn}\nolimits}
\def\dn{\mathop{\rm dn}\nolimits}
\def\am{\mathop{\rm am}\nolimits}
\def\abbr#1{{\rm #1}}

\newcommand{\be}{\begin{equation}}
\newcommand{\ee}{\end{equation}}
\newcommand{\bea}{\begin{eqnarray}}
\newcommand{\eea}{\end{eqnarray}}
\newcommand{\rd}{{\rm d}}
\newcommand{\p}{\partial}

\newcommand{\thalf}{{\textstyle\frac{1}{2}}}

\usepackage{graphicx}
\usepackage{dcolumn}
\usepackage{bm}

\begin{document}


\title{Electrostatics of Edge States of Quantum Hall Systems with Constrictions: 
Metal--Insulator Transition Tuned by External Gates}

\author{Emiliano Papa$^1$ and Tilo Stroh$^2$} 

 \address{$^1$Department of Physics, The University of Virginia, Charlottesville, 
VA 22904-4714\\
 $^2$Department of Physics, University of Siegen, 57068 Siegen, Germany}
\date{\today}

\begin{abstract}

The nature of a metal--insulator transition tuned by
external gates in quantum Hall (QH) systems with point constrictions at integer bulk filling, as reported
in recent experiments \lbrack S. Rodaro {\it et al.}, Phys. Rev. Lett. {\bf 97}, 046801 (2005)\rbrack, is addressed.   
We are particularly concerned here with the insulating behavior---the phenomena
of backscattering enhancement  induced at high gate voltages.   
Electrostatics calculations for QH systems with split gates performed here show that 
observations are not a consequence of interedge interactions near the point contact. 
We attribute the phenomena of backscattering enhancement 
to a splitting of the integer edge into conducting and insulating stripes, which enable the 
occurrence of the more relevant backscattering processes of
fractionally charged 
quasiparticles at the point contact.
For the values of the parameters used in the experiments we find that the conducting channels are  
widely separated by the insulating stripes 
and that their presence alters significantly the low-energy dynamics of the edges.
Interchannel impurity scattering does not influence strongly the tunneling exponents as they are 
found to be irrelevant processes at low energies. 
Exponents of backscattering at the point contact are unaffected by interchannel Coulomb interactions 
since all channels have the same chirality of propagation.
\end{abstract}

\maketitle

\section{Introduction} 

An electron gas in two spatial dimensions (2DEG) in the presence of a strong magnetic field at particular values of the filling 
fraction (ratio of the number of electrons to flux quanta penetrating the sample) forms a charge gap in the bulk and gapless 
excitations at the edges creating there one-dimensional (1D) chiral metallic states.  These states have attracted a great amount of interest as 
an arena for studying the exotic properties of the Luttinger liquid (LL)
model \cite{chang_0}.

Even though remarkable theoretical progress has been achieved in
understanding the properties of Luttinger liquids,\cite{gogolin,voit}
in the experimental front open problems remain.\cite{grayson,levitov}
For instance, 
questions have been raised from transport experiments, across a gate-created
constriction in quantum Hall (QH) systems,  
by Roddaro {\it et al.}\ \cite{pellegrini}.  The measurements in Ref.~\onlinecite{pellegrini} observe, in QH samples at integer bulk filling, 
a metal--insulator transition induced by varying the voltage of the metallic gates ($V_\abbr{g}$) that define the constriction. 
At low $|V_\abbr{g}|$ the system is 
metallic (backscattering is suppressed) across the point contact and becomes insulating (backscattering is enhanced) as 
$|V_\abbr{g}|$ is increased.  At high $|V_\abbr{g}|$ the observed current
across the constriction has a bias-voltage evolution that would be reminiscent of a bulk-filling fraction $\nu=1/3$. 

\begin{figure} 
\begin{center}
\includegraphics[width=0.8\columnwidth]{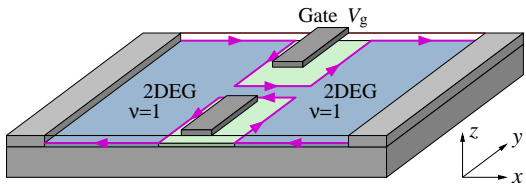} 
\end{center}
\caption{ 
Schematic illustration of a quantum Hall bar with a split-gate
constriction similar to those used in experiments of Roddaro {\it et al.}
\protect\cite{pellegrini}.  The electrostatic top gates bring counter-propagating edges in close proximity
to enhance interedge backscattering. More details of actual devices are given in Ref.~\onlinecite{pellegrini}.}
\label{schematic_illustration_0}
\end{figure}

These types of tunneling experiments intend to probe predictions of the LL theory such as the power-law dependence of the 
interedge tunneling correlation functions and the related power-law suppression of the densities of states.  
In QH systems points of enhanced backscattering are created by bringing counter-propagating edges in close 
proximity by either placing metallic split gates over the top of the 2D electron gas, 
like the ones illustrated schematically in
Fig.~\ref{schematic_illustration_0}, \cite{pellegrini}
or by allowing for breaks in 
barriers like the ones obtained by cleaved edge overgrowth \cite{kang} 
(in which the interedge spacing  
is on the order of a magnetic length enabling additional electron
correlations along the barrier).
The theoretical groundwork for single-point
tunneling phenomena in Luttinger liquids was laid by Kane and Fisher \cite{kane} by using
a renormalization-group scheme.
At low energies a change in nature of transport across a backscattering point 
depending on the value of the Luttinger liquid parameter was predicted:
perfect transmission for attractive interactions across the sample
and a perfect reflection at the point contact for repulsive interactions.
These predictions were later confirmed  
by Fendley, Ludwig and Saleur \cite{fendley} 
using the Bethe-ansatz technique which established the exact flow between
the Dirichlet, ${\cal T} \to \infty$, the strong-tunneling stable fixed
point, and the Neumann, 
weak-tunneling unstable fixed point.  Here ${\cal T}$ is the tunneling amplitude.
Contrary to the experiments of Ref.~\onlinecite{pellegrini}
the LL theory predicts a rather
uninteresting  voltage-independent top--bottom tunneling conductance ($G\sim V^{2\nu -2}$) 
at filling $\nu=1$, typical for Fermi liquids. 
It is the intention of this paper to examine the disagreement between theory and experiment.

The low $|V_{\abbr g}|$ behavior of the differential tunneling conductance at a constriction in relation with
experiments \cite{pellegrini_0} and \cite{heiblum} was examined in \cite{papa2}.
It was concluded that repulsions between electrons on opposite sides of the gates 
suppress the quasiparticle (QP) backscattering at the constriction---thus leading to the metallic behavior observed 
at low $|V_{\abbr g}|$.
The idea there was that edges on opposite sides of the 
Hall bar must be regarded as counter-propagating channels of a LL formed along the gates, rather than as the left
and right portions of the top or bottom chiral channels of the overall Hall bar. 
Repulsive interactions between the chiral channels of such a LL are expected to enhance the left--right 
electron tunneling. 

The reason for the occurrence of the opposite phenomena:
the enhancement of backscattering at the constriction observed at
high values of $|V_{\abbr g}|$, in systems similar to those illustrated
in Fig.~\ref{schematic_illustration_0}, was analyzed in
Ref.~\onlinecite{Papa_Stroh_PRL2006}. 
In this paper, we explain in significant detail the results that were
discussed in Ref.~\onlinecite{Papa_Stroh_PRL2006} and elaborate
on other experimental implications in connection with the proposed model.
In analogy to the low $|V_{\abbr g}|$ case, one might be tempted to attribute the experimental 
observations at high $|V_{\abbr g}|$ to the possibility of sign-reversed interactions 
across the barrier due to the presence of image charges in the metallic gates.
To study the nature of interactions between edges on opposite sides
of the gates, we 
allow for different potentials on the respective 2DEGs, and 
study the change of position of one edge upon a change of potential on the other 2DEG.
We find in fact that electron interedge interactions remain weakly repulsive at any values of these voltages.
The other venue that we pursue in this paper is the investigation of the effects of the edge-reconstruction phenomena that 
take place in wide and smooth edges inevitably 
created at high gate voltages, and caused by the presence of magnetic fields in QH samples. 
In the absence of a magnetic field we find that the width of the QH
edge increases linearly on $|V_{\rm g}|$, with only logarithmic $|V_{\abbr g}|$ corrections, favoring the need for an
electrostatic treatment of such smooth edges.  The charge distribution along the edges in the presence of magnetic fields changes.
With the introduction of charge energy gaps at points of fractional filling, the edge 
breaks into a sequence of compressible (CS) and incompressible (IS) stripes, the latter exhibiting 
the fractional QH effect, QHE, (FQHE).  
Quite generally such stripes form in any situation where at zero magnetic fields a charge-density 
gradient is present, as noted in \cite{efros,glazman,chklovskii}. 
The ISs are strongly electron-correlated states 
and exhibit insulating properties as opposed to CSs 
which behave as metal stripes
(of constant potential in the area they occupy). 
The charge density remains with a gradient in the compressible ones and 
the low energy physics of the edge is described by their fluctuations \cite{vignale}. 
Large part of this paper is dedicated to the calculation of positions and dimensions of the 
CSs and ISs.
The values of $|V_\abbr{g}|$ of Ref.~\onlinecite{pellegrini} for the geometry of 
these systems  
turn out to be such that for bulk $\nu=1$ filling the ratios of 
ISs to the CSs 
are of the order
$\approx 1$ and $\approx 0.5$, 
for the outermost and the next to the outermost ones, respectively. 
The presence of these wide insulating stripes 
enables the backscattering of fractionally $(e/3)$ charged quasiparticles whose tunneling exponent would correspond to an effective
 bulk $\nu=1/3$ filling.
The interchannel interactions do not influence this exponent. The latter has its physical origin in 
the known fact that interchannel interactions cannot break the universality of the Hall conductance 
in same-chirality multichannel edges like the ones at  higher integer Landau level filling  Hall systems.  

This paper is organized as follows. In Sect.\ II we examine the functional
 dependence of the edge width and edge position as a function
 of the voltage of the gates.  Subsequently, in Sect.\ III we examine the
 nature of interedge interactions in the presence of metallic gates.
 In Sect.\ IV we find the geometric dimensions of the compressible and
 incompressible stripes that form at the edges of the 2DEGs in the presence 
 of a magnetic field.
In Sect.\ IV we find the geometric dimensions of the compressible and
incompressible stripes that form at the edges of the 2DEGs in the presence
of a magnetic field.
In Sect.\ V  we discuss implications of the reconstructed edge in transport 
and finite-frequency noise experiments. 
We close the paper with our conclusions.

\section{Structure of Edges in Split Hall Bar Systems in the Absence of Magnetic Fields}
\label{sect:sym-case}
\label{sect:edg-struct-B-abs}

In this section based purely on electrostatic arguments we study the charge distribution and edge structure and 
edge position of a 2D electron gas in the presence of a confining potential provided by electrostatic split gates. 
We first examine     
the case of the absence of a magnetic field. The geometry of the model is represented
schematically in Fig.~\ref{schematic_illustration_0}.  The 2DEG resides on the
interface of two semiconductors, the $xy$ plane ($z=0$) of Fig.~\ref{schematic_illustration_0}.
Two metallic gates are deposited over the top a distance $\approx 10\,l_B$ from the 2DEGs' plane along 
the $y$ axis of Fig.~\ref{schematic_illustration_0}.
$l_B = |\hbar c/Be|^{1/2}$ 
is the magnetic length, $\approx 100$\,\AA,
for a magnetic field $B=6.5$\,T that will be introduced later. The electron charge here is $-e$ and $c$ is the light constant.
We assume here nevertheless that the gates and the 2DEGs are coplanar. The gates, of lateral
extension $x \in (-a,a)$, are used to create the constriction to enhance the interedge electron tunneling amplitude.
As the gate voltage is increased, the electrons under
the gate are repelled leaving behind positively charged dielectric stripes parallel to the $y$ 
axis of Fig.~\ref{schematic_illustration_0},  extended in the intervals $a < |x| < b$. 
In equilibrium, there is no net electric field on the 2DEGs' plane, and the 2DEGs can be treated 
as metallic components with constant potential. 
The equilibrium position will be reached at the points where the horizontal component
of the electric field created by the charged dielectric stripes and the one created by
the metallic components (the 2DEGs and the gate) sum to zero. 
At these points we require 
\bea
\label{eq:lim-x+b+}
 \lim_{x\to \pm b\mp 0}\Bigl\{E_x^{\rm str.}(x,z=0)+E_x^{\rm el.}(x,z=0) \Bigr\}
\rightarrow 0 \quad . 
\eea
Each of those fields is singular at the sharp edges but in equilibrium they 
cancel exactly.

\begin{figure}
\begin{center}
\includegraphics[width=0.55\columnwidth]{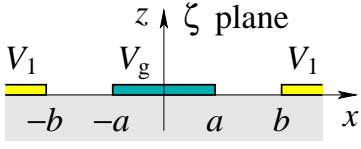} 
\end{center}
\caption{ 
Schematic illustration of split gates on the $y=0$ plane at high gate voltages.
For high values of $|V_{\abbr g}|$ the metallic gate and the electron gases can be assumed to be 
coplanar.  The potential of the 2DEGs on the sides of the gate are taken to be equal to $V_1$. 
The positively charged stripes of the host semiconductor extend in the depleted regions $a < |x| < b$.
}
\label{schematic_illustration}
\end{figure}

Semiconductor structures in experiments \cite{pellegrini} were fabricated from AlGaAs/GaAs single 
heterojunctions of charge density
$n_0 \approx 10^{11}$\,cm$^{-2}$ 
and dielectric constant $\epsilon=12.6$.
In the depleted region the charge density is defined by the positive
background charge of the donor semiconductor, $en_0$. 
In the 2DEGs the electron charge density increases gradually away from the boundaries
(located at $\pm b$), reaching values of $-en_0$ in the bulk. We are 
interested here in studying the case of high gate voltage, which 
justifies the assumption of putting the 2DEGs, the positively charged dielectric stripes,
and the metallic gates all in the same plane, $z=0$ of Fig.~\ref{schematic_illustration}.

We assume here that the system is translational invariant along the gates. 
The problem then becomes effectively two-dimensional. 
A knowledge of the electric-field configuration in space leads to the knowledge of the electric charge distribution
in space and thus to the knowledge of the structure of the edge which is the purpose of this section.
We proceed in the following with the calculation separately of the electric fields 
created by the metallic components and the electric field created by the charged dielectric stripes.
The potential on the plane $z=0$ fulfills the boundary conditions
\bea
  &&\phi(x,z=0) = \left\{
  \begin{array}{r@{\quad:\quad}l}  V_\abbr{g} &
\label{bc1}
    |x|<a  \quad , \\[1ex]
  V_1 & |x| > b
  \quad ,       
  \end{array}\right.
  \\
  &&\left.\frac{\p \phi(x,z)}{\p z}\right|_{z\rightarrow 0^-} = 
  \frac{4\pi e n_0}{\epsilon} \quad, \quad 
  a < |x| < b \quad .
\label{bc2}
\eea
The latter is justified by the fact that the dielectric medium has a
large dielectric constant of $\epsilon=12.6$ [and $4\pi/(\epsilon+1)\approx 4\pi/\epsilon$].

 We solve in the following the Laplace equation in the half-plane $z<0$ using the boundary conditions 
(\ref{bc1}) and (\ref{bc2}).
We separate the solution 
into a sum
$\phi(x,z) = \phi^{\rm el.}(x,z) + \phi^{\rm str.}(x,z)$ 
of harmonic functions \cite{glazman} that satisfy separately the conditions
\bea
  &&\phi^{\rm el.}(x,z=0) = \left\{
  \begin{array}{r@{\quad:\quad}l}  V_\abbr{g} & |x|<a  \quad , \\[1ex]
                 V_1 - A(a,b) & |x| > b \quad ,
  \end{array}\right.
  \label{met_components1}
 \\
  &&\left.\frac{\p \phi^{\rm el.}(x,z)}{\p  z}\right|_{z\rightarrow 0^-} =
  0 \quad, \quad
  a < |x| < b \quad ,
  \label{E_z_metals}
\eea
for the metallic components. The latter equation reflects the fact
that the $z$ component of the electric field in the depleted regions
$a < |x| < b $, for symmetry reasons, is zero. On the other hand,
for the potential created by the dielectric stripes,
$\phi^{\rm str.}(x,z)$, we have
\bea
  &&\phi^{\rm str.}(x,z=0) = \left\{
  \begin{array}{r@{\quad:\quad}l} 0 &  |x| < a  \quad ,
  \\[1ex]
  A(a,b) & |x| > b \quad,
  \end{array}\right.
  \label{outside_interv}
 \\
  \label{inside_interv}
  &&\left.\frac{\p \phi^{\rm str.}(x,z)}{\p  z}\right|_{z\rightarrow 0^-} =
  \tau(x) \quad  , \quad a < |x| < b\,  \quad .
\eea
In these equations $\tau(x)$ is the charge of the dielectric stripes,
\[
  \tau(x) =\frac{4 \pi e n_0}{\epsilon}
  \left\{ \begin{array}{r@{\quad:\quad}l}
  0 & |x| < a \quad\mbox{or}\quad |x|>b \\[1ex]
  1 & a < |x| < b
  \end{array} \right.
  \quad,
\]
and $A(a,b)$ is a constant
specific to our calculations and will be defined below. 
The boundary conditions fulfilled by $\phi^{\rm str.}(x,z)$ consist in the knowledge
of the vertical component of the electric field in the depleted region and
in the fact that the potential in  the 2DEGs and gate regions is constant.

To determine $\phi^{\rm str.}(x,z)$ we generalize here an idea given by Glazman and Larkin \cite{Glm+Lkn}
by representing the potential $\phi^{\rm str.}(x,z)$ as the imaginary part of an
analytic function $F(\zeta)$, where $\zeta=x+\imu z$. 
By imposing the boundary conditions (\ref{outside_interv}), (\ref{inside_interv}) on the derivative
$\rd F/\rd \zeta$ with an appropriate choice of an auxiliary function $D(\zeta)$, Eq.~(\ref{f_function02}), we can 
relate it to another function, $f(\zeta)=\imu D(\zeta)(\rd F/\rd\zeta)$,
whose imaginary part we would know everywhere along the $x$ axis.
By taking the opposite steps (see Appendix \ref{App_A}) we determine
\bea
  \label{stripes_pot}
  \lefteqn{\phi^{\rm str.}(x,z)=
  \tau b \RE\left\{\bE(k')- 
  E\left(\psi,k'\right)
  \vphantom{\frac{A}{B}}
  \right.}
  \\[0ex]\nonumber
  && {} - \frac{1+k^2}{2}\left[\bK\left(k'\right)
  \right.
  -\left. \left. F\left(\psi,k'\right)\right]
  + \imu \left(\frac{\zeta}{b} - k \right)\right\}
  \quad ,
\eea
where 
$\psi = \arcsin[\frac{1 - (\zeta/b)^2}{1-a^2/b^2}]^{1/2}$
and $\tau = {4\pi e n_0/\epsilon}$
(more details are given in Appendix \ref{App_A}).
Here and in the following $\bK=\bK(k)$, $\bE=\bE(k)$,
$\bK'=\bK(k')$, $\bE'=\bE(k')$ are the complete 
(complementary complete) Jacobi
elliptic integrals of the $1^{\rm st}$ and $2^{\rm nd}$ kind 
of modulus $k=a/b$ and $k'=(1-k^2)^{1/2}$, \cite{gradshteyn} respectively.
The potential constant $A(a,b)$ introduced in (\ref{met_components1}),(\ref{outside_interv}) 
is given in (\ref{A(ab)}).

The potential $\phi^{\rm el.}(x,z)$ created by the metallic components 
with the boundary conditions (\ref{met_components1}) and (\ref{E_z_metals})
can be determined by using $\pi/2$ folding conformal transformations 
at the points $\pm a$, $\pm b$,
mapping the upper  half-space of Fig.~\ref{schematic_illustration} to a rectangle. 
To fulfill 
(\ref{E_z_metals}),
corresponding in the transformed space to a vanishing horizontal component of 
electric field on the boxes' vertical sides,
the space inside the box has to be part of a larger space,
obtained by successive iterative reflections of the box around its vertical sides, 
reaching eventually to an infinitely long capacitor whose plates are at potentials $V_{\rm g}$ 
(lower plate) and $V_1 - A$ (top plate). 
The potential resulting from such a transformation \cite{conformal}
is obtained from 
\bea
  \label{metal_field}
  \Phi(\zeta) = \frac{{V}_1 - A - V_\abbr{g}}{\bK(\sqrt{1-a^2/b^2})} \sn^{-1}
  \left(\frac{\zeta}{a},k\right) + \imu V_\abbr{g}
\eea
with $\phi^{\rm el.}=\IM\Phi$.
The electric field created by the metallic components is obtained from 
$ E_x^{\rm el.}=  - {\rm Im}[ {\rd \Phi}/{\rd \zeta} ] |_{z=0}$, whereas the 
one of the dielectric stripes is obtained from
$E^{\rm str.}_x = - {\rm Im}[\rd F/\rd \zeta]|_{z=0}$, cf.\ 
(\ref{eq:F-der-zeta}). 
Both of these fields exhibit singularities at the edge boundaries.
To reach mechanical equilibrium at the boundary positions, 
one requires that
 \begin{equation}
V_1 - V_\abbr{g} = \frac{4\pi e n_0 b}{\epsilon}\Bigl[\bK(\sqrt{1-a^2/b^2}) - \bE(\sqrt{1-a^2/b^2})\Bigr]
\;.
  \label{eq:position_equation}
\end{equation}
This equation\cite{larkin_shikin} defines the positions $\pm b$ of the edges of the 2DEGs.

For small values of the dimensionless constant 
$\kappa:=(V_1 - V_\abbr{g})/\tau a$, 
with the expansion series 
for small values of the argument in complete elliptic
integrals,\cite{gradshteyn} one gets
the linear relation $b/a=1+2\kappa/\pi$. 
For the limit of large $\kappa$, for which $b-a\gg a$, 
we make use of  the logarithmic series \cite{gradshteyn}   
and get the asymptotic large $|V_{\rm g}|$ dependence of $b$:
\be
  b=\frac{\kappa a}{ \ln (4\kappa/{\eun})-\ln\ln (4\kappa/{\eun})}
  \quad , \quad
  \kappa = (V_1 - V_\abbr{g})/\tau a \;.
\ee
(Here the Euler number $\eun$ is not to be confused with the
elementary charge $e$.)
The distance of the edge position increases linearly on $V_1-V_{\abbr g}$ with only logarithmic corrections.
A plot of the dependence of the 
edge position as a function of the potential difference $|V_1-V_{\abbr g}|$
obtained by the numerical solution of the transcendental 
equation (\ref{eq:position_equation}) is shown in Appendix \ref{app:calc-metpot-gen} (Fig.~\ref{schematic_illustration_2}).

The charge density distribution on the 2DEGs' regions, the $z=0$ plane
at $|x|>b$, can be found similarly, by the knowledge of the $z$-component electric field.
In the $z=0$ plane the stripes give a contribution of
$E^{\rm str.}= 
- \left. \RE\left[{\rd F}/{\rd \zeta}\right] \right|_{z=0}$.
The derivative of $F$, with the appropriate behavior at $x\to \pm \infty$
is given by 
${\rd F/\rd \zeta} =-\imu  \tau G^{\rm str.}(\zeta) + \tau$, where from Eq.~(\ref{eq:F-der-zeta}) we have 
$G^{\rm str.}(\zeta) = {[\zeta^2 - (a^2+b^2)/2]}/D(\zeta)$.
In the regions $|x| > b$ we find
\be
  N(x) = n_0(G^{\rm str.}(x)-1)-(V_1 - A  - V_\abbr{g})G^{\rm el.}(x) ,
\label{total_charge}
\ee 
which is the total charge density $eN(x) = -en(x) + en_0$, 
a superposition of the positively charged dielectric background of 
density $en_0$ and the 2DEGs in $|x|>b$ on top---carrying charge density $-en(x)$.
In Eq.~(\ref{total_charge}) the short notation $ G^{\rm el.}(\zeta) =  b/[\bK' D(\zeta)]$ is used.
The profile of the edge of the electron liquid 
follows to be
\bea
  n(x) = n_0 \left( \frac{x^2-b^2}{x^2-a^2} \right)^{1/2} \quad , \quad
  |x| > b \quad .
  \label{profile_eq}
  \label{eq:profile_eq}
\eea
At large distances from the edge the charge density reaches the expected bulk value of $n_0$.
The width of the edges increases as $(b-a)$, becoming
smoother at higher $|V_{\abbr g}|$. 
Numerical estimates for the edge position in devices used in Ref.~\onlinecite{pellegrini}, for 
$V_1 - V_\abbr{g}=1\,{\rm V}$ and  
$a=2500 \,{\rm \AA}$ is $b=4964 \,{\rm \AA}$. 
The distance of the gate to the 2DEG is $\approx 1000 \,{\rm \AA}$.
This justifies our assumption of taking the 2DEG and the gate as being coplanar.

\section{Nature of interactions}
\label{section:nature}

In this section we study the influence of the metallic gates on the
electron interactions between edges on opposite sides of the gates.
Previous work \cite{papa2} has found that QP or electron backscattering in the constriction can be 
enhanced or suppressed depending on the sign of the interedge interactions in the constriction. 
Attractive interactions, for instance, would effectively lower
the filling fraction in the Hall bar 
and enhance the backscattering of the emerging QP at the point contact.
In general cases the influence on the electron interactions of nearby metallic gates
is hard to estimate. The geometry examined here, however, is such that it allows
for exact calculations.
Intuitively one would expect that the presence of image charges in the gate, halfway the between edges,
to weaken the interedge Coulomb repulsions or even reverse their sign.

To study these effects we 
allow for different voltages on the 2DEG subsystems.
In the following we calculate the positions of the edges as a function
of the voltages and look at the ratios of the change of the position
of one edge upon the change of voltage on each of the 2DEGs.
Due to the voltage difference between the 2DEGs, their distance from
the metallic gate should be different. We assume in general the edge
boundaries be positioned at $\bar b$ and $b$ whereas the gate be extended
within the interval $\bar a < x < a$, with $\bar a - \bar b \neq b - a $.

\subsection{Field of Asymmetric Dielectric Stripes} 
\label{sect:noi-fld-str}

\begin{figure*}
\begin{center}
\unitlength=1.2mm
\begin{picture}(86,30)
\put(0,16){\includegraphics[width=36.7\unitlength]{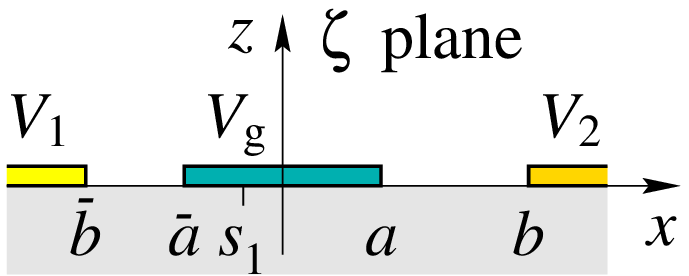}}
\put(0,0){\includegraphics[width=36\unitlength]{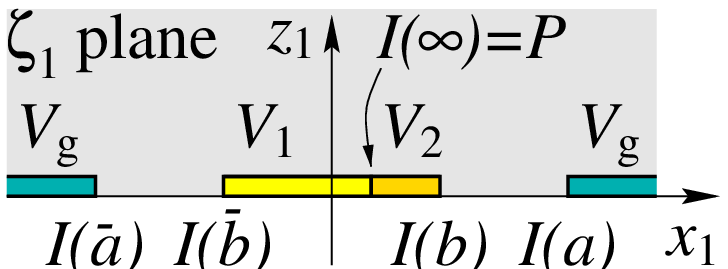}}
\put(38,-2){\includegraphics[width=48\unitlength]{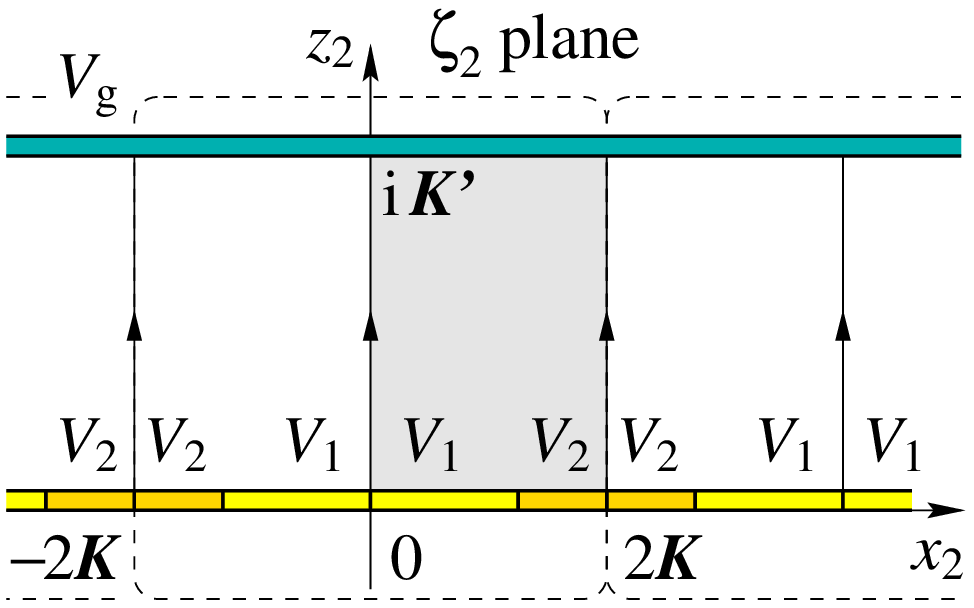}}
\end{picture}\nolinebreak
\end{center}
\caption{ 
Upper left: original potential configuration in
space---metallic gate and 2DEG potential.
Since $V_1\neq V_2$ the distance
of the 2DEGs
from the gate is different. Lower left: after
inversion around a point $s_1$ the 2DEGs' distances from the gate
are equal.
Right: forming a box configuration (shaded) by mapping
the upper $\zeta_1$ half-plane, which in turn
 maps the lower half-plane of the original $\zeta$ plane (\ref{z_to_z2}).
Application of the Schwarz mirroring principle about the vertical borders
of the shaded box leads to a periodic structure along the $x_2$ axis.
}
\label{inverted_system}
\end{figure*}

Here we calculate the electric field and the potential configuration created by the positively charged
asymmetric stripes and its constant potential surroundings. 
The dielectric stripes extend in the intervals $(\bar b,\bar a)$ and
$(a,b)$ as shown in Fig.~\ref{inverted_system} (upper left).
The boundary conditions for the dielectric stripes now have to be modified to
\bea
  \label{eq:bndc-gen-out2}
  \!\!&\!&\phi^{\rm str.}(x,z=0) =  \left\{\begin{array}{r@{\quad:\quad}l}
     0  & \bar a < x < a  \quad , \\[0.5ex]
     A  &   x < \bar b \quad, \\[0.5ex]
     B  &    x > b \quad,
  \end{array}\right.
  \\*
  \label{eq:bndc-gen-in2}
  \!\!&\!&\left.\frac{\p \phi^{\rm str.}(x,z)}{\p  z}\right|_{z\rightarrow 0^-} =
  \tau(x) \;, \; \bar b < x < \bar a \; ,  \; a < x < b \; ,
  \qquad
\eea
where $\tau(x) = 4\pi e n_0/\epsilon$ 
and $A$, $B$ are constant potentials that will be defined below.

We impose the boundary conditions (\ref{eq:bndc-gen-out2}),
(\ref{eq:bndc-gen-in2}) on $\rd F/\rd \zeta$, where
$\phi^{\rm str.}(x,z) = \IM F(\zeta)$ and $\zeta = x + \imu z$.  We introduce as before the auxiliary function
$D(\zeta)$ (Ref.\ \onlinecite{D_F})
to build $f(\zeta)=\imu D(\zeta) \rd F/\rd \zeta$
such that $f(\zeta)$ takes imaginary values everywhere on the $x$ axis.
Using the Schwarz equality (\ref{eq:Schwarz}), we get in analogy
to (\ref{eq:f=i(D+F)}):
$f(\zeta) = \tau[\imu D(\zeta) + {\cal F}(\zeta,\bar b,\bar a,a,b)]$.
The function ${\cal F}(\zeta,\bar b,\bar a,a,b)$
again must be chosen with the correct
behavior at infinity, $f(\zeta)\rightarrow 0$
as $|\zeta|\rightarrow \infty$, resulting in ${\cal F}(\zeta,\bar b,\bar a,a,b)$ given in \cite{D_F}
for constant $\tau$.
The knowledge of ${\cal F}(\zeta,\bar b,\bar a,a,b)$ immediately leads to the
knowledge of the electric field in the depleted regions, $z=0$,
$\bar b < x < \bar a$,  $a < x < b$,
\be
  E_x= \tau \,  \frac{{\cal F}(x,\bar b,\bar a,a,b)}{D(x)}
  \quad .
  \label{eq:stripes-fld-x-gen}
\ee
The electric field created by the stripes, as one would expect,
is singular at every edge (including the edge boundaries $\bar b$ and $b$ of the 2DEGs).

The potential created by the dielectric stripes, $\phi^{\rm str.}(\zeta)$, after integration
of $f(\zeta)/D(\zeta)$ (Ref.\ \onlinecite{byrd}) 
is found to be given by the expression (\ref{eq:asy-F-gen}).
The constants $A$ and $B$ are defined as potential differences
$A= \phi^{\rm str.}(\bar b,0)-\phi^{\rm str.}(\bar a,0)$ and 
$B = \phi^{\rm str.}(b,0)-\phi^{\rm str.}(a,0)$.
From (\ref{eq:asy-F-gen}) the expression for $B$ turns out to be 
\be
 \label{eq:asy-k-const-gen}
 B=
 \frac{\tau}{4}S_k^-
 \left(\frac{b-\bar b}{a-\bar a}+\frac{a-\bar a}{b-\bar b}\right)\bK(k')
 -\frac{\tau}{2}S_k^+\bE(k') \;,
\ee
where the modulus of the elliptic functions is [transformed from $\tilde k$ of (\ref{eq:asy-F-gen}) to] 
$k=S_k^-/S_k^+$ 
with $S_k^\pm=\left[{(b-\bar a)(a-\bar b)}\right]^{1/2} \allowbreak\pm\allowbreak \left[{(b-a)(\bar a-\bar b)}\right]^{1/2}$.
The expression for $B$ is invariant under $\bar b \leftrightarrow b$,
$\bar a \leftrightarrow a$,
implying that $A=B$. This constant represents the value of the
potential at large distances from the gates. 

\subsection{Field of Gates and 2D Electron Gases} 
\label{sect:noi-fld-2DEG}

 As in the previous sections, we assume that the metallic gates and the 2DEGs are coplanar.
The boundary conditions for the potential of the metallic components now have to be modified to
\bea
  \label{outside_interv2}
  \!\!&&\phi^{\rm el.}(x,z=0) =  \left\{\begin{array}{l@{\quad:\quad}l}
     V_\abbr{g} &  \bar a< x < a  \quad , \\[0.5ex]
     \tilde{V}_1 = V_1 - A & x < \bar b \quad, \\[0.5ex]
     \tilde{V}_2 = V_2 - B & x > b \quad,
  \end{array}\right.
  \\*
  \label{inside_interv2}
  \!\!&&
  \left.\frac{\p \phi^{\rm el.}(x,z)}{\p  z}\right|_{z\rightarrow 0^-} = 0
  \;: \; \bar b < x < \bar a \; ,  \; a < x < b \; .
\eea

 The potential difference of the 2DEGs renders the $\pi/2$ folding transformations,
used previously, not directly applicable.
We circumvent this limitation and determine the field configuration in space by employing the following steps.

First, make an inversion transformation around a point $s$ on the
$x$ axis under which the points $a_i$, $i=1,\ldots, 4$ corresponding to $\bar b ,\ldots, b$ of the $\zeta=x+\imu z$
plane map to points $I(a_i)$ as follows,
\be
  \label{inv_1}
  \zeta_1=\frac{R^2}{\zeta-s}+P \quad , \quad I(a_i)=\frac{R^2}{a_i-s}+P \quad .
\ee
$R$, the radius of inversion, introduces an overall length scale. The potential field is scale invariant
and independent of $R$, however.
$P$ merely shifts the origin of the $\zeta_1$ plane, 
which we take to be the middle point between $I(\bar b)$ and $I(b)$, 
see Fig.~\ref{inverted_system} (lower left).
Finally we choose $s$ such that the distances between the 2DEG edges and the
metallic gate in $\zeta_1$ space  be equal, $I(\bar a)-I(\bar b)= I(b)-I(a)$.
This condition is resolved at two points, $s_{1,2}$, \cite{s_12} from which only $s_1$ is used below.
The transformation in (\ref{inv_1}) 
inverts the space bringing the points $\pm \imu \infty$
to $\mp\imu 0$ and $\pm \infty$ to $\pm 0$. 
In $\zeta_1$ space the gate lies in the outer regions and the dielectric 
stripes around the origin enclosing the 2DEGs.
The 2DEG subsystems of potentials $\tilde{V}_1$ and $\tilde{V}_2$ are located, respectively, between $I(\infty)$ and 
$I(\bar b)$ for $\tilde{V}_1$, and between $I(b)$ and $I(\infty)$ for $\tilde{V}_2$.

Second, we perform another conformal transformation mapping the new
configuration to a box [$\pi/2$ folding transformations at the points
$I(a_i)$, $i=1,\ldots,4$].
This brings us in the $\zeta_2$ plane,
\be
  \zeta_2  = 
      {\rm sn}^{-1}\left(\frac{\zeta_1}{I(b)},\frac{I(b)}{I(a)}\right)
  + \bK\left(\frac{I(b)}{I(a)}\right)\quad ,
  \label{sn_transf}
  \label{z_to_z2}
\ee
where we have selected the multiplicative and the additive constants
be 1 and $\bK$, respectively, as shown also in
the right panel of Fig.~\ref{inverted_system}. 
The modulus of the Jacobi elliptic integral $\bK$ is $I(b)/I(a)$, which equals $k=S_k^-/S_k^+$ of the Jacobi integrals contained in 
the expression for $B$ of the previous section.

To fulfill the boundary conditions (\ref{inside_interv2}) of vanishing electric field along $\hat{z}$ 
in the depleted regions in $\zeta$ space, the shaded box of Fig.~\ref{inverted_system} (right) in $\zeta_2$ space
has to be part of the periodic configuration shown on the same figure, 
in which the electric field
vanishes on the vertical borders of the shaded box. 

The expression for the potential field in the new space can be calculated analytically
to the close form (see Appendix \ref{app:calc-metpot-gen}),
\begin{eqnarray}
 \label{close_form}
  \lefteqn{
 \phi^{\rm el.}(\zeta_2,{\bar \zeta}_2)
  = \tilde V_2-\left(V_\abbr{g}-\tilde V_2
  -\frac{\tilde V_1-\tilde V_2}{2\bK}C\right)
    \frac{\RE\{\imu \zeta_2\}}{\bK'}
 }  \nonumber\\[2mm] && {}
+\frac{\tilde V_1-\tilde V_2}{\pi}
    \RE\left\{\imu\ln\frac{\vartheta_1\left(\frac{\pi}{4\bK}(\zeta_2+C),
    \euz^{-\frac{\pi\bK'}{2\bK}}\right)}
   {\vartheta_1\left(\frac{\pi}{4\bK}(\zeta_2-C),
   \euz^{-\frac{\pi\bK'}{2\bK}}\right)}
    \right\}
\; . 
\qquad
\end{eqnarray}

Here $\vartheta_1$ is an elliptic theta function \cite{gradshteyn} and $C$ is the reflection of the 
$\pm \infty$ points of $\zeta$ space in $\zeta_2$ space, $C=\zeta_2(I(\pm\infty))$. 
This field configuration can now be mapped onto the original plane.

Now we have the necessary ingredients to
proceed on determining the functional dependence of the 2DEGs'
edge positions $\bar b$, $b$
on $V_1$ and $V_2$.

\subsection{Edge-state Position as a Function of Potentials and
the Influence of Metallic Gate on Interedge Interactions}

\label{sect:edge_st_position}

Since the edges of the 2DEGs are mobile, their equilibrium position is reached at points where the $x$ component of the electric field 
in the $z=0$ plane of the overall system vanishes. Once again we use the mechanical equilibrium equation similar to Eq.~(\ref{eq:lim-x+b+}): 
\bea
\left. E_x(x,z=0)\right|_{x=\bar b} = 0 \quad , \quad \left.E_x(x,z=0)\right|_{x= b} = 0 \quad ,
\label{eq:lim-x-a+,d-}
\eea
and try to turn these conditions into a system of equations for the edge boundaries $\bar b$, $b$.
The $x$-component electric field at $x\to\bar b$ and $x\to b$ in the $z=0$
plane of both, the dielectric stripes, as well as the metallic components
in principle are known. 
The calculations in this section consist primarily of a simplification of the expressions for the electric field of the metallic components at $z=0$.
We seek a reformulation of $ E^{\rm el.}_{y_2}$ from the elliptic $\vartheta_1$ functions, of modulus $\exp\{-\pi \bK'/2\bK\}$, 
to elliptic integrals of the same modulus as the ones of the expression for
the potential $B$ of Eq.~(\ref{eq:asy-k-const-gen}) that is present in 
the potentials of the capacitor plates. For the sake of clarity we give some details here, but leave the rest 
to Appendix \ref{app:Ey2-eqs}. 

First, for the electric field of the metallic components  
the relationship
between $E^{\rm el.}_x$ and $E^{\rm el.}_{y_2}$ at the points
$\zeta_2 = 2n\bK+\imu y_2$ [corresponding to
points in the intervals $x\in (\bar b,\bar a)$ and $x\in (a,b)$  and $z=0$ in $\zeta$ space] is needed.
Using the 
equalities (\ref{inv_1}), (\ref{sn_transf}) one can show that at the points
$\zeta_2=2n\bK+\imu y_2$, where the field does not have a
horizontal component in $\zeta_2$ space, the Jacobian of the transformation $\zeta \to \zeta_2$ takes a simple form, and
hence
\be
E^{\rm el.}_x(x,z=0) = \left. E^{\rm el.}_{y_2}(\zeta_2,\bar{\zeta}_2) \frac{\p \zeta_2}{\imu \p \zeta_1}\frac{\p \zeta_1}{\p \zeta} 
\right|_{x_2=2n\bK}
\quad .
\label{eq:E-coord-tr(2,0)}
\ee
For the coordinate transformation (\ref{eq:E-coord-tr(2,0)}), 
with the explicit form of $s_1$ from 
Ref.\ \onlinecite{s_12}, one obtains:
\bea
J(2\to 0)=  - \imu \frac{\partial \zeta_2}{\partial \zeta_1} \frac{\partial \zeta_1}{\partial \zeta} 
   &=& -\frac{1}{2}\frac{S_k^+}{D(\zeta)} \;. 
  \label{Jacobian}
\eea
The Jacobian of the transformation to the original space provides
the expected square-root singularities
for the $x$ component of the electric field at the
edge boundary points $\bar b$ and $b$.
The sign of the electric field is determined by the factor $D(\zeta)$
along the line $\zeta=x-\imu 0$.

Second, the vertical component of the electric field in $\zeta_2$ space
is given by
\begin{widetext}
\bea
  \label{eq:Ey2-by-theta}
E^{\rm el.}_{y_2}(\zeta_2,\bar{\zeta}_2) =
  -\frac{\partial\phi^{\rm el.}(\zeta_2,\bar{\zeta}_2)}{\partial y_2} =
  -\tilde V_0 \frac{1}{\bK'}
+\frac{\tilde V_1-\tilde V_2}{4\bK}\RE\left\{
  \frac{\vartheta_1'\left(\frac{\pi}{4\bK}(\zeta_2+C),
  \euz^{-\frac{\pi\bK'}{2\bK}}\right)}
  {\vartheta_1\left(\frac{\pi}{4\bK}(\zeta_2+C),
  \euz^{-\frac{\pi\bK'}{2\bK}}\right)}-(C\to-C)\right\}
\quad ,
\eea
\end{widetext}
where the prime in $\vartheta_1$ denotes the derivative with respect to the first argument of $\vartheta_1$ and
$\tilde V_0=V_\abbr{g}-\tilde V_2-(\tilde V_1-\tilde V_2)C/(2\bK)$ was introduced. 

\begin{figure}
\begin{center}
\includegraphics[width=0.65\columnwidth]{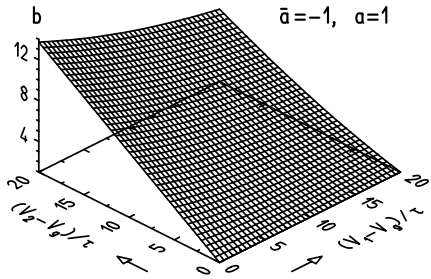}
\end{center}
\caption{ Position of the edge, $b$, on the right of the metallic gate as a function of $V_1-V_\abbr{g}$ 
and $V_2-V_\abbr{g}$.  At negative gate voltage $b$ increases past the edge of the metallic gate. 
The dependence for fixed $V_1$ is linear on $V_2-V_\abbr{g}$. For $V_1$ equal to $V_2$ the distance 
of the edge positions from the gate is linear in agreement with the one shown in Fig.~\ref{schematic_illustration_2}. }
\label{edge_d}
\end{figure}

The conditions (\ref{eq:lim-x-a+,d-}) involve the fields $E^{\rm el.}_{y_2}$ at $\zeta_2=0,2\bK$,
(\ref{E_y2_0}), (\ref{E_y2_2K}). With (\ref{eq:asy-k-const-gen_new}) to
(\ref{eq:k-sq-dn-terms}) and by using $V_i$ instead of $\tilde V_i$ and
$V_0=\tilde V_0-B$ one gets for their symmetric and
antisymmetric combinations,
\bea
 \label{eq:lim-x-0}
   &&
   E^{\rm el.}_{y_2}(2\bK)-E^{\rm el.}_{y_2}(0) =
   -2 \frac{(V_1-V_2)}{\pi} \frac{(b-\bar b)}{S_k^+} \;, \\[0mm]
   &&
   E^{\rm el.}_{y_2}(2\bK)+E^{\rm el.}_{y_2}(0) =
   -2\frac{B+V_0}{\bK'}
   -2\frac{V_1-V_2}{\pi}\frac{l}{S_k^+}
   \;, \qquad
 \label{eq:lim-x-2K}
\eea
where the length $l$ is given by Ref.~\onlinecite{length_l}.
On the other hand, the field created by the charged dielectric stripes is obtained from (\ref{eq:stripes-fld-x-gen}).
Inserting these field expressions and (\ref{eq:asy-k-const-gen}) into the
difference and sum of fields obtained by multiplying (\ref{eq:lim-x-a+,d-})
by $D(x)$, we get\vadjust{\penalty-200}
\bea
  \label{position_1}
  &&\frac{b+\bar b}{2}=\frac{a+\bar a}{2}-\frac{V_1-V_2}{\pi\tau}
  \;,
  \\[1mm]
  \label{position_2}
  &&  0 = \frac{\tau}{2}(b-\bar b)^2
  -\frac{\tau}{2} \frac{\left({S_k^+}\right)^2\bE'}{\bK'}
  +\frac{S_k^+ V_0}{\bK'}+\frac{V_1-V_2}{\pi}l
  \qquad
\eea
with $\left({S_k^+}\right)^2 \le (b-\bar b)^2$ and $\bE'\le\bK'$.
Equations (\ref{position_1}) and (\ref{position_2}) define a complete system from which the dependence
of the edge positions $\bar b$ and $b$ of the 2DEGs on the voltages
of their respective potentials can be found.  We have solved these equations self-consistently numerically
and have plotted results for $b(V_1,V_2)$ for fixed $V_\abbr{g}$ in
Fig.~\ref{edge_d}. 
The plot for  $\bar b(V_1,V_2)$ is equivalent to $b(V_1,V_2)$ under the exchange $V_1 \leftrightarrow V_2$, 
$x \leftrightarrow -x$. In the limit when the voltages on the 2DEG subsystems are equal,
$V_1=V_2$, we have $\bar a = -a$, $\bar b = -b$ and get $S_k^+ \ =2b$. 
The equations for the edge positions in this limit simplify, yielding an equation consistent with the earlier result 
[Eq.~(\ref{eq:position_equation})] 
that was derived slightly differently in Sect.\ \ref{sect:edg-struct-B-abs}.

The quantity of importance in our calculations here is the sign of $\p_{V_1} b/\p_{V_2} b$.
By increasing $V_1$, as can be seen in Fig.~\ref{edge_d}, $|\bar b|$ increases, as does $b$ upon the increase of $V_2$.
The graph in Fig.~\ref{fig:d-pd-V1,V2} shows that upon the increase of
$V_1$ the 2DEG edge $b$, on the other side of the gate, approaches the gate 
as the other edge $\bar b$ goes away from the gate. This signifies that interedge interactions remain repulsive in 
nature  at any values of $V_1$ and $V_2$. At low values of $V_i-V_{\abbr g}$, $i=1,2$, instead $\p_{V_1} b/\p_{V_2} b \to 0$, 
signifying the tendency for the complete screening of interedge interactions from the gate. 
We conclude that the explanation of  enhancement of
top--bottom backscattering (the insulating behavior across
the constriction) observed in the experiments Ref.~\onlinecite{pellegrini} 
has to be related with other phenomena taking place 
at high gate voltages.  We elaborate on this in the following section.
\begin{figure}
\begin{center}
\includegraphics[width=0.65\columnwidth]{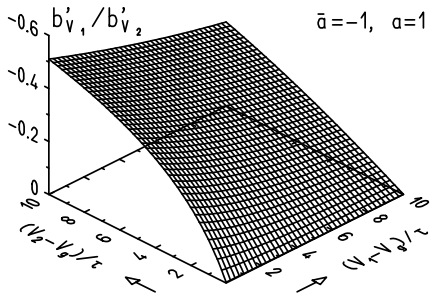}
\end{center}
\caption{$\partial_{V_1} b/ \partial_{V_2} b$, the ratio of the differential
dependences of the edge position on the potentials of the 2DEGs on the sides of the gate.  
By increasing $V_1$ of the 2DEG subsystem on the left of the gate of
Fig.~\ref{schematic_illustration}, the edge on the right-hand side
approaches the gate, just as the other one on the left goes away from it.
This signifies that interedge interactions in the presence of the metallic
gate remain repulsive in nature.  }
\label{fig:d-pd-V1,V2}
\end{figure}

\section{Structure of the Edge in the Presence of Magnetic Fields}

\label{sect:edg-struct-B-pres}

\begin{figure}
\begin{center}
\unitlength=1mm
\includegraphics[width=0.65\columnwidth]{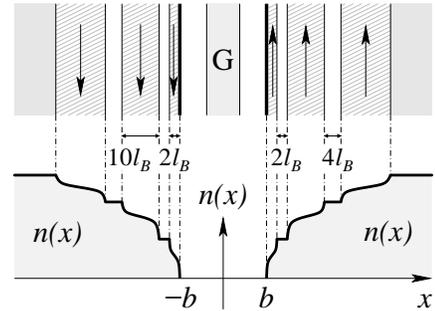}
\end{center}
\caption{
Structure of the edge of a 2DEG surrounding the
split gate, in the presence of a strong perpendicular magnetic field
in QH systems with constrictions.
Shaded areas notify current-carrying CSs,
separated by insulating ISs. 
For half a gate width of $a=5000 \,{\rm \AA}$ the geometrical dimensions of the stripes are
$180 \,{\rm \AA}$ for the first CS, $180 \,{\rm \AA}$ for the $1/3$ IS, $1040 \,{\rm \AA}$
for the second CS, and $377 \,{\rm \AA}$ for the $2/3$ IS, respectively. 
}
\label{schematic_illustration_3}
\end{figure}

The effect of strong perpendicular magnetic fields on the structure of
the edge of a 2DEG is known from earlier work of Refs.\ 
\onlinecite{efros,glazman,chklovskii}, and \onlinecite{chang}. 
In the edge of a bulk integer $\nu=1$, around points 
of $f=p/(2mp \pm 1)$ filling, for $p$ and $m$ integers, a hierarchy of
FQH energy gaps exists. The filling fraction as usual is defined by $\nu = n_0/2\pi l_B^2$
as the number of electrons per flux quantum. 
With the introduction of the gaps, the potential in the 2DEGs is not constant throughout 
as in Sect.\ \ref{sect:edg-struct-B-abs}.
As we move in the $x$ direction away from the edge boundaries,  
  the potential varies  in a discontinuous manner, jumping in value by $V_f$ 
each time we cross the location of a FQHE gap corresponding to filling $f$. 
The system therefore will tend to lower its energy if some electrons are relocated from higher to
lower energies forming this way fractional-filling ISs   
around points of $f=p/(2mp \pm 1)$, each of which exhibits the FQHE.
  The energy profile across the edge varies  in a step-like manner, being constant in the 
compressible metallic regions and increasing across the incompressible insulating stripes. 
Therefore application of a
strong magnetic field will in principle break the edge of the 2DEG confined 
by a smooth external potential into a sequence of
CSs and ISs.   
The widest ISs    
will be the ones with the largest gaps [see also Eq.~(\ref{eq:incstr-wd-asy}) below]
and these will be the ones realized in practice. In the case
of integer fillings the role of the FQH energy gaps will be played by the
much larger Landau level splitting. As a result the integer edges 
will lead to wider and more stable incompressible stripes.
The positions of the incompressible stripes correspond to fractional
fillings and in general follow equipotential lines parallel
to the external gate (Fig.~\ref{schematic_illustration_3}).
 
The largest charge gaps in the FQHE are known to occur at $f=1/3$ and
$2/3$ fractions\cite{chang} 
and in  the following we calculate their widths. 
In general the position of a fractionally charged stripe of filling fraction
$f$ 
can be found by substituting $n(x_{f\pm})=f n_0$ in Eq.~(\ref{profile_eq}) and solving
it for $x_{f\pm}$. Such substitution leads to the $x_{f\pm}$ positions
\bea
  x_{f\pm} = \pm b \left( \frac{1 - k^2 f^2 }{1-f^2}  \right)^{1/2}\quad .
\eea
Based on the estimate of Sect.\ \ref{sect:edg-struct-B-abs} for
$b=1.986\,a$
at 
$V_1-V_\abbr{g} = 0.674\,{\rm V}$
and $B=6.5\:$Tesla, 
$a=25\,l_B$, we find
\bea
     x_{1/3\pm}=\pm 2.08\, a \quad , \quad x_{2/3\pm} = \pm 2.51\, a \quad .
\eea

In this section from the knowledge of $V_f$'s we calculate the widths of the sequence of CSs and ISs on the edge.
 It is assumed that the metallic gates and the 2DEGs are coplanar.
The electrostatics problem in the presence of a magnetic field
is:\cite{glazman}
\bea
  \label{b.c.}
  \!\!\!&&\phi'(x,z=0) = \left\{\begin{array}{r@{\;\,:\;\,}l} 
       V_\abbr{g}  & \;\; \; \quad     |x| < a   \quad , \\[1ex]
       V_1         & b < |x| < a' \quad , \\[1ex]
       V_1 + V_{f}   & \;\; \;  \quad    |x| > b'  \quad ,
  \end{array}\right.
  \\[1ex]
  \!\!\!&&\left. \frac{\p \phi'(x,z)}{\p z} \right|_{z\to 0^-} \! = 
  \left\{\begin{array}{r@{\;\,:\;\,}l}
    \frac{4\pi e n_0}{\epsilon} & a < |x|< b  \; , \\[1ex]
    \frac{2\pi e }{\epsilon} (n_0 - n_{f})  &
      a' < |x| <  b'  \; .
  \end{array}\right.
  \quad
  \label{b.c.2}
\eea
In (\ref{b.c.2}) a factor of two is missing when compared with (\ref{bc2}). This is due to the 
fact that the ISs are narrow when compared with the 2DEG distance from the surface of the semiconductor, 
and the electric field surrounding the ISs can be thought of as concentrated  within the semiconductor.
In (\ref{b.c.}) $V_f$ is the voltage difference on the sides of the IS
of filling fraction $f$ that arises due to a discontinuity on the value
of the chemical potential of elctrons across the IS. $V_f$ is given by
$V_f=\Delta \mu_f /e$, where for integer filling IS,
$\Delta \mu=\hbar \omega_{\abbr{c}}$, whereas for fractional fillings
$f=p/q$, where $q$ is an odd integer, it is given by
$\Delta \mu_f= q c_f e^2/\epsilon l_B$. 
The constants $c_f$ are $c_{1/3}=c_{2/3}=0.03$
(Ref.\ \onlinecite{chang} and references therein).
The quantity $c_f e^2/\epsilon l_B$, 
the charge gap, is the energy necessary to create a pair of
(particle--hole) quasiparticles of charge $e/q$. 
The biggest charge gaps are known to occur at $f=1/3$ and 2/3 fillings.
In the following, therefore, due to the smallness of the charge gaps, the
ISs at other fractions will not be considered. 
For $f=1/3$
and $f=2/3$ we have $V_f=1.026 \,{\rm mV}$.  
As in Ref.~\onlinecite{glazman} the solution of the Laplace equation  in the half-plane $z<0$ can be found as a sum
$\phi' = \phi'^{\rm el.} + \phi'^{\rm str.}+ \phi_0$ of harmonic functions
$\phi'^{\rm el.}$, $\phi'^{\rm str.}$ and the zero-magnetic-field solution $\phi_0$.
We model the boundary conditions for $\phi'^{\rm el.}$ as the ones of the
metallic 2DEGs of Sect.\ \ref{sect:edg-struct-B-abs} 
and those of $\phi'^{\rm str.}$ as
the ones belonging to the corresponding charged dielectric stripes. 
The boundary conditions for $\phi'^{\rm el.}$ then read
\bea
\label{el_field_new}
  \phi'^{\rm el.} (x,z=0) = \left\{\begin{array}{r@{\quad:\quad}l}
    0  & |x|< a'  \, , \\[1ex]
    \tilde{V}_1' & |x| > b' \, ,
  \end{array}\right.
  \\[1ex]
  \left. \frac{\p \phi'^{\rm el.} (x,z)}{\p z} \right|_{z\to 0^-} =  0 \quad , \quad
  a' < |x| < b' \quad ,
\eea
where $\tilde{V}_1' =  V_f - A(a',b')$. In analogy to the boundary
conditions (\ref{outside_interv}), (\ref{inside_interv}) for $\phi^{\rm str.}(x,z)$
the potential $\phi'^{\rm str.}(x,z=0)$ of the charged stripes formed at the
edge profile fulfills
\bea
  \phi'^{\rm str.} (x,z=0) = \left\{\begin{array}{r@{\quad:\quad}l}
    0 & |x|< a'  \, , \\[1ex]
    A(a',b')  & |x| > b'  \, ,
  \end{array}\right.
  \label{eq:IS-plane-pot}
  \\[1ex]
  \left. \frac{\p \phi'^{\rm str.} (x,z)}{\p z} \right|_{z\to 0^-}
\hspace{-4mm}  =  \frac{2\pi e}{\epsilon}[n(x) - n_f] \; , \;
  a' < |x| < b' \; .
  \label{diel_stripes_new}
\eea
The only difference with the previous case is the $x$ dependence of the
charge density of the stripes extended in $a' < |x| < b'$ on the
sides of $x_f$. 
Qualitatively, a presence of the potential difference $V_f$ 
on the sides of $x_f$ will tend to push away the sides, and the remaining stripes will 
provide the traction for the electrostatic equilibrium at the points $a'$ and $b'$. 
The fact that the stripes are now charged both positively on the left of $x_f$ and negatively 
on the right of $x_f$ will result in a reduction of the traction at small $V_f$ as compared to the 
case of uniform positively charged stripes provided by the dielectric of
Sect.\ \ref{sect:noi-fld-str},
leading to a rapid opening of ISs at small $V_f$. Nevertheless, as $V_f$ 
increases the positive part of the charged stripe will become strong,
leading to 
an eventual slowed-down increase
of the width of the ISs as a function of $V_f$ at higher values of $V_f$.

\begin{figure}
\unitlength=1mm
\begin{center}
\includegraphics[width=0.6\columnwidth]{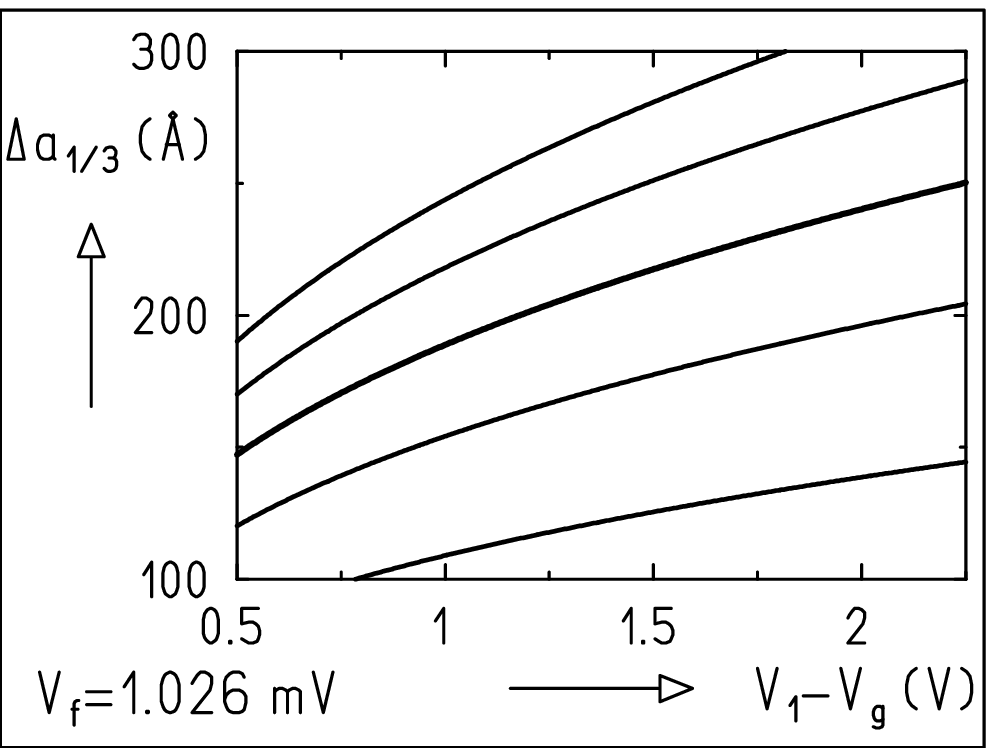}
\bigskip\par
\includegraphics[width=0.6\columnwidth]{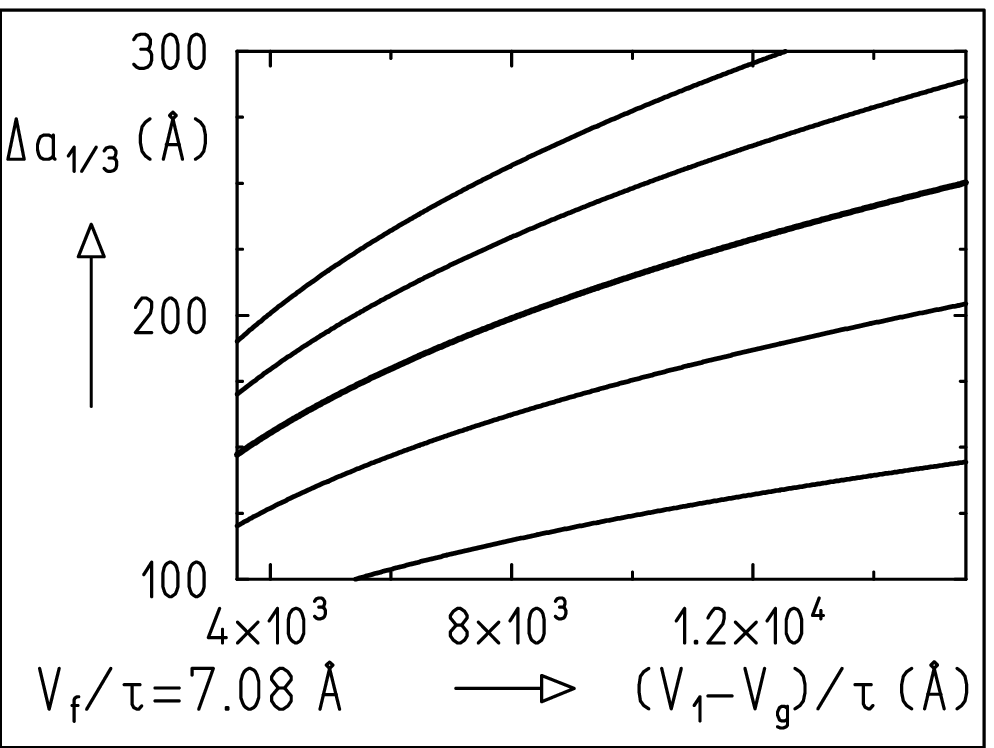}
\end{center}
\caption{
Plots of width of incompressible stripes of $1/3$ filling $\Delta a_{1/3}$
in \AA\ as a function of
$V_1-V_\abbr{g}$ and $V_f$ for the second-order approximation
and the values $V_f\to (i/3)V_f$, $i=1,\ldots,5$, from bottom to top are shown
(in Ref.\ \protect\onlinecite{pellegrini}, $V_1-V_\abbr{g}=0.674\,$V 
and $V_f=1.026 \,{\rm mV}$).
The gate width is taken to be $2a$ with $a=25 \, l_{B}$. 
}
\label{fig:stripe_widths-2d}
\end{figure}

Since the values of the charge gaps are relatively small when compared with 
$\left|V_1-V_\abbr{g}\right|$, 
$V_f/\left|V_1-V_\abbr{g}\right|\approx 10^{-3}$ 
in Ref.\ \onlinecite{pellegrini}, 
the charge density in the region of ISs, ${\cal N}(x) = n(x) - n_f$, 
can be expanded around the points $x_{f\pm}$.
The even function $n(x)$ is approximated by one even polynomial
$\sum_{i=0}^N \tilde n_i^{(N)} x^{2i}$ matching $n(x)$ up to the
$N$th derivative at $x_{f\pm}$. In the following 
the notation $\tau(x) = (2\pi e/\epsilon) {\cal N}(x)$ is used.
In electrostatic equilibrium the positions of the IS  
edges are obtained by requiring the electric field created
by the metallic components extended outside the intervals
$a' < |x| < b'$ and the charged stripes, be zero at all points 
$\pm a',\pm b'$.

In principle the calculations for the electric fields $E^{\rm el.}(x)$
and $E^{\rm str.}(x)$ satisfying the boundary conditions
(\ref{el_field_new})--(\ref{diel_stripes_new}) for 
$\phi'^{\rm el.}(x)$ and $\phi'^{\rm str.}(x)$,
are similar to the ones of Sect.\ \ref{sect:edg-struct-B-abs}. 
The only difference arises from the fact 
that the charge density of the stripes has now an $x$-dependent profile. In the following we 
describe briefly the calculations for $E^{\rm str.}(x)$, leaving details of the calculation
in Appendix \ref{appx:pol-chrg-dist}. 
By representing the potential $\phi^{\rm str.}$ as the imaginary part of an
analytic function $F$, and using the boundary conditions for
its derivative, we relate this to the function $f$, $f(\zeta)=\imu D(\zeta)(\rd F/\rd\zeta)$,
$\zeta = x+ \imu z$, which in turn after using Eq.~(\ref{eq:Schwarz}) takes the form
$f(\zeta)=\tau[\imu D(\zeta) + {\cal F}(\zeta)]$, where
the differences with the previous case come primarily in the form of ${\cal F}(\zeta)$.
${\cal F}(\zeta)$ will depend on the form of the $D(\zeta)$ as well as on $\rd F/\rd \zeta$, i.e.,
on the function $\tau(x)=\sum_n \tau_n x^n$ of (\ref{diel_stripes_new}), 
where $n$ will be the order of our approximation.
The general form of ${\cal F}(\zeta)=\sum_n{\cal F}_n$ results form the knowledge of 
\begin{equation}
  {\cal F}_n(\zeta,a',b')=
   \tau_n\zeta^{n+2}
\hspace{-1mm} 
\sum_{k=0}^{\left[n/2+1\right]}
\hspace{-2mm}
   C_k^{-\frac{1}{2}}\left({\textstyle \frac{a'^2+b'^2}{2a'b'}}\right)
   \left(a'b'/\zeta^2\right)^k
  \hspace{-1mm},
\end{equation}
where $C_k^{-\frac{1}{2}}(t)$ is a Gegenbauer polynomial, 
and $[\cdot]$ represents the integer part
of a real number. $\phi^{\rm str.}$ is found straightforwardly by integrating $f(\zeta)/D(\zeta)$.

In the following we will calculate the cases of $n=2$ and $n=4$ of the approximation.

\subsubsection{Second-Order Expansion}

\label{sect:esBp-1st-ord}

Here we take the lowest order of approximation for $n(x)$ of
(\ref{profile_eq}), i.e., $N=1$,
\bea
  n(x)\approx n_f + \frac{n_f'}{2x_f}\Bigl(x^2-x_f^2\Bigr)
\quad ,
\label{eq:n(x)-1st-ord}
\eea
with $n'_f\equiv \left.\p_x n(x)\right|_{x_f}$.
The $x$ dependence of the charge density of the stripes extended in
the intervals $a'<|x|<b'$ 
will change the results slightly
as compared with the ones of Sect.\ \ref{sect:edg-struct-B-abs}. 
The result in this case for the previously introduced $F$ function,
whose imaginary part gives the potential, is obtained by the integration
of (\ref{eq:F-der-zeta}) and given in (\ref{eq:F^1}).
For $A$ in Eq.~(\ref{eq:IS-plane-pot}) we get 
 \begin{eqnarray}
 \label{A_1}
 \label{eq:A_1}
   \lefteqn{
   A^{(1)}(a',b')=\tau_0^{(1)} b'
   \left({\textstyle\frac{1}{2}}(1+{k}^2)\bK'
   - \bE'\right)} \\* \nonumber
   &&{}+
   {\textstyle\frac{1}{6}}\tau_2^{(1)} b'^3
   \left[\left({\textstyle\frac{3}{4}}\bar{k}'^4 + 2\bar{k}^2 \right)\bK'
   -(1+\bar{k}^2)\bE'\right] \quad,
 \end{eqnarray}
with $\tau_0^{(1)}=-x_f\tau'/2$,
$\tau_2^{(1)}={\tau'}/{2x_f}$, $\tau'=2\pi e n_f'/\epsilon$,
and $\bar k=a'/b'$.
 As in Sect.\ \ref{sect:sym-case} the value of the potential constant $A(a',b')$ is specific for our calculations.
Requiring for the total value of the electric field at both edges of
the stripe to vanish, we obtain two equations, one of which reads
\be
  \tau'(b'^2-a'^2)\left(\frac{x_f}{4}-
  \frac{a'^2+3b'^2}{16x_f}\right)
  =\frac{V_f-A^{(1)}(a',b')}{\bK'}b' \;,
  \label{eq:incstr-eqs-1st}
\ee
and for the other one $a'$ and $b'$ are interchanged on the left-hand side.
Their numerical solution would lead to the stripe width $\Delta a_f=b'-a'$.
From the difference of (\ref{eq:incstr-eqs-1st}) with its
left-hand side ($a' \leftrightarrow b'$) interchanged 
a constriction for the edge boundaries of the ISs  
is gained,
\be
  a'^2+b'^2=2x_f^2 \quad .
  \label{eq:incstr-1st-eq1}
\ee
$x_f$ is away from the middle point between
$a'$ and $b'$ shifted toward
$b'$, a feature that becomes more pronounced for wider ISs. 
Taking the sum of (\ref{eq:incstr-eqs-1st})
with its left-hand side ($a' \leftrightarrow b'$) interchanged 
using also $A^{(1)}(a',b')$ from
(\ref{A_1}) we obtain as second equation for the position of the
stripe boundaries 
$a'$ and $b'$: 
\bea
  \left(1-\frac{\bar k'^2}{2}\right)\bE' -\bar k^2\bK'
  &=& 3\left(1 - \frac{\bar k'^2}{2}\right)^{\frac{3}{2}} \frac{V_f}{\tau'x_f^2} \,.
  \label{eq:incstr-1st-eq2}
\eea
At the small values of $V_f$ considered here
the above equality therefore will be
resolved at values $\bar k\to 1$ ($b'-a'\ll a'$).
In this limit using expansions of $\bE'$ and $\bK'$
for small $\bar k'$, \cite{dwight} for $\bar k$ we get 
$\bar k\approx 1- (2/x_f)(2V_f/\pi \tau')^{1/2}$,
which gives
$ 
 \Delta a_{1/3} \approx  \left[{4 \epsilon V_f}/{\pi^2 e
  \p_x n(x_{1/3}) }\right]^{{1}/{2}}
$.
The stripes
will be wider for bigger gaps and 
around points with smallest charge gradient. 
Therefore the ones located further from the edge boundary (inner) ISs will be wider than the ones located closer to the boundary (outer ones).

\begin{figure}
\begin{center}
\includegraphics[width=0.6\columnwidth]{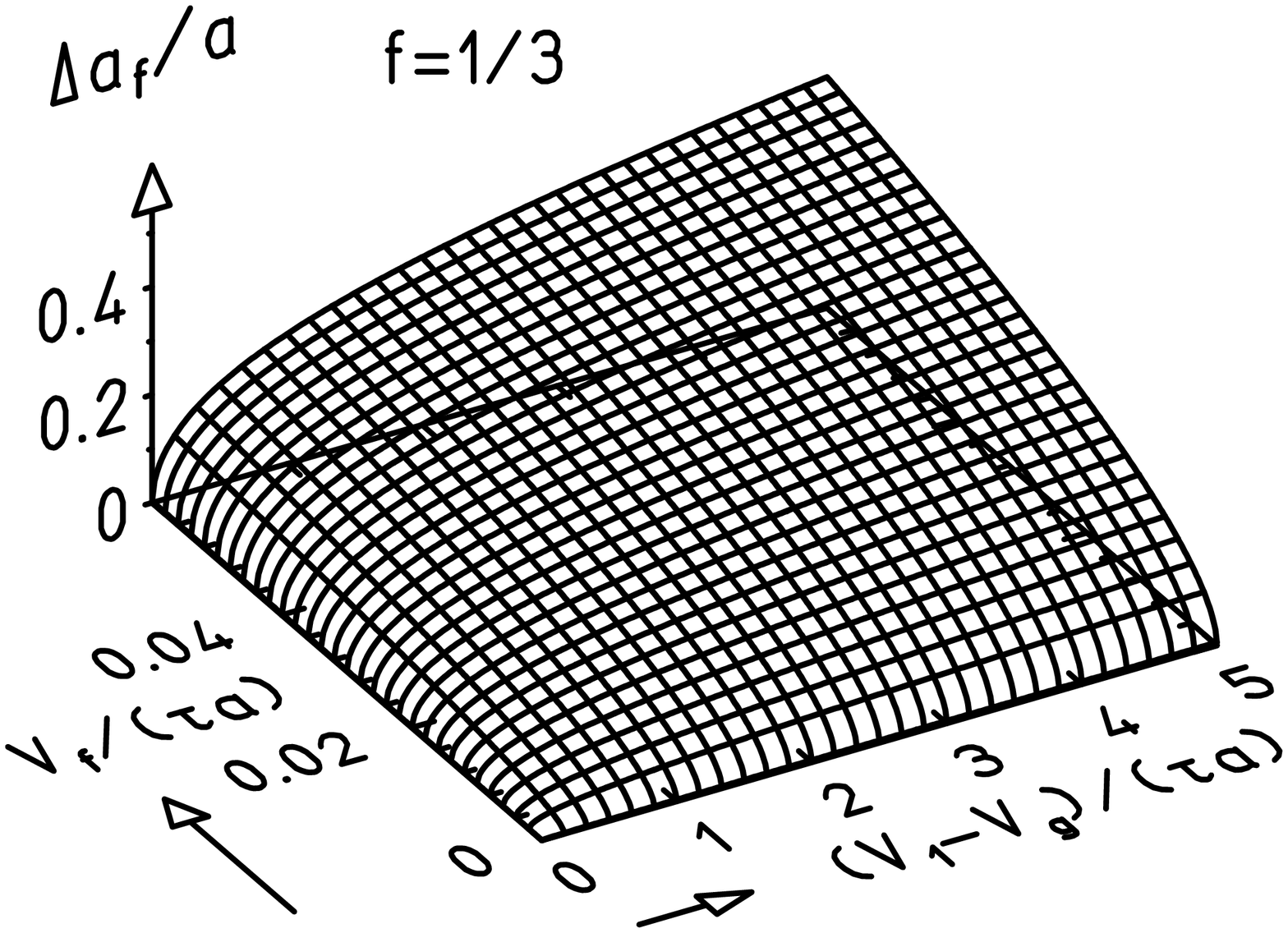}
\bigskip\par
\includegraphics[width=0.6\columnwidth]{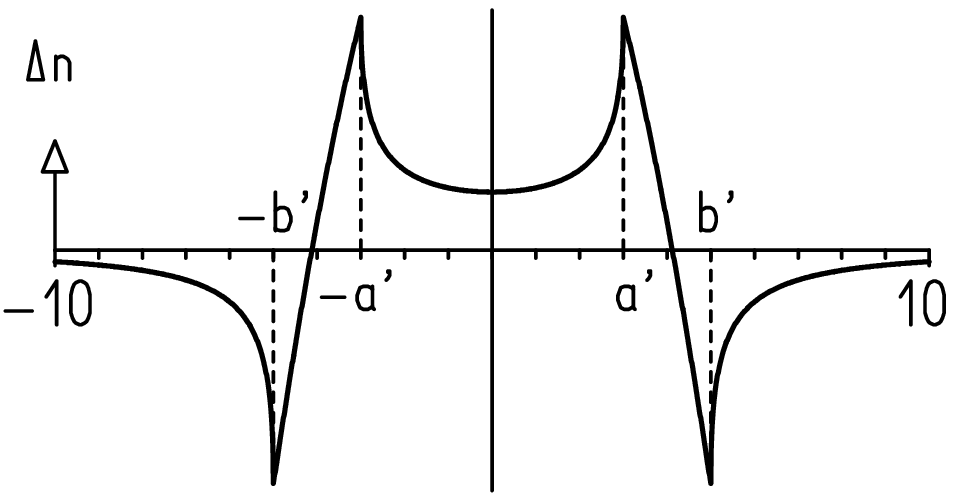}
\end{center}
\caption{
Top: three-dimensional plot for stripe widths $\Delta a_f$ for $f=1/3$,
measured in units of half of the gate width, $a$,
taken usually in Ref.~\onlinecite{pellegrini} to be 
$a=2500$\,\AA--$5000$\,\AA.
Bottom: Schematic plot of the additional electron charge caused by the presence
of a strong magnetic field on the edge profile of a QH state wide
enough to support ISs.
$a'= 3a$, $b'= 5a$ are chosen as the positions of the boundaries of the
$1/3$ IS on the right of the gate in this illustration.
The point $a'<x_f<b'$, where $n(x_f)=fn_0$,
is defined in the text. 
}
\label{fig:stripe_widths-1st-3d}
\label{charge_imbalance}
\end{figure}

Because asymptotically $x_f$ increases linearly with $b$, $n'_f$
decreases accordingly, and so the term $1/(\tau'x_f^2)$ from
(\ref{eq:incstr-1st-eq2}) decreases inverse linearly.
Therefore its solution has to be taken in the limit
$\bar k\to 1$ ($b'-a'\ll a'$). 
The widths of ISs, $\Delta a_f=b'-a'$, 
turn out to be\cite{factor} 
\bea
  \Delta a_f \approx \frac{\epsilon}{\pi e n_0}
  \left(\frac{1}{\pi} \frac{f V_f}{(1-f^2)^{\frac{3}{2}}}
  \frac{(V_1-V_{\abbr g})}{\ln \left[4(V_1-V_{\abbr g})/\tau a \eun \right]}
  \right)^{\frac{1}{2}}
  \;,
  \label{eq:incstr-wd-asy}
\eea
proportional to $(V_1-V_{\abbr g})^{1/2}$, up to logarithmic corrections. 
The dependence of the IS widths is also proportional to $V_f^{1/2}$,
in sharp contrast to the 
previous case, of the uniformly charged stripes. Such a behavior has
its origin in the fact 
that the stripes have a charge profile in $x$ direction. Plots of the
dependence of $\Delta a_f$
on $V_1-V_{\abbr g}$  and $V_f$, obtained by the numerical solution of
(\ref{eq:incstr-1st-eq1}) and (\ref{eq:incstr-1st-eq2}),
are shown in Figs.~\ref{fig:stripe_widths-2d} and
\ref{fig:stripe_widths-1st-3d} (top).

For the narrower outer ISs we find a width of
$\approx 165$\,\AA--$180$\,\AA. 
The wider inner $2/3$ channels, of size
$\approx 340$\,\AA--$\approx 380$\,\AA,
are
$\approx 800$--1000\,\AA\
away.

For the additional charge, induced by the magnetic field in forming
the ISs 
at $|x|>b$, we find
\[
  \Delta n = \frac{n_f'}{2x_f}\left(\frac{a'^2+b'^2}{2}-x^2+ \left\{
  \begin{array}{lcl}
    \left|D'(x)\right| &:& |x|>b' \\[0.5ex]
    -\left|D'(x)\right| &:& |x|<a'
  \end{array} \right\}\right) ,
\]
where $D'(x)$ is obtained by $D(x)$ of Sect.\ \ref{sect:edg-struct-B-abs} by the
substitution $a \to a'$ and $b\to b'$.
A plot, on a magnified scale, of $\Delta n$ along the system of two edges is shown
in Fig.~\ref{charge_imbalance} (bottom).

\subsubsection{Fourth-Order Expansion}

\begin{figure}
\begin{center}
\includegraphics[width=0.6\columnwidth]{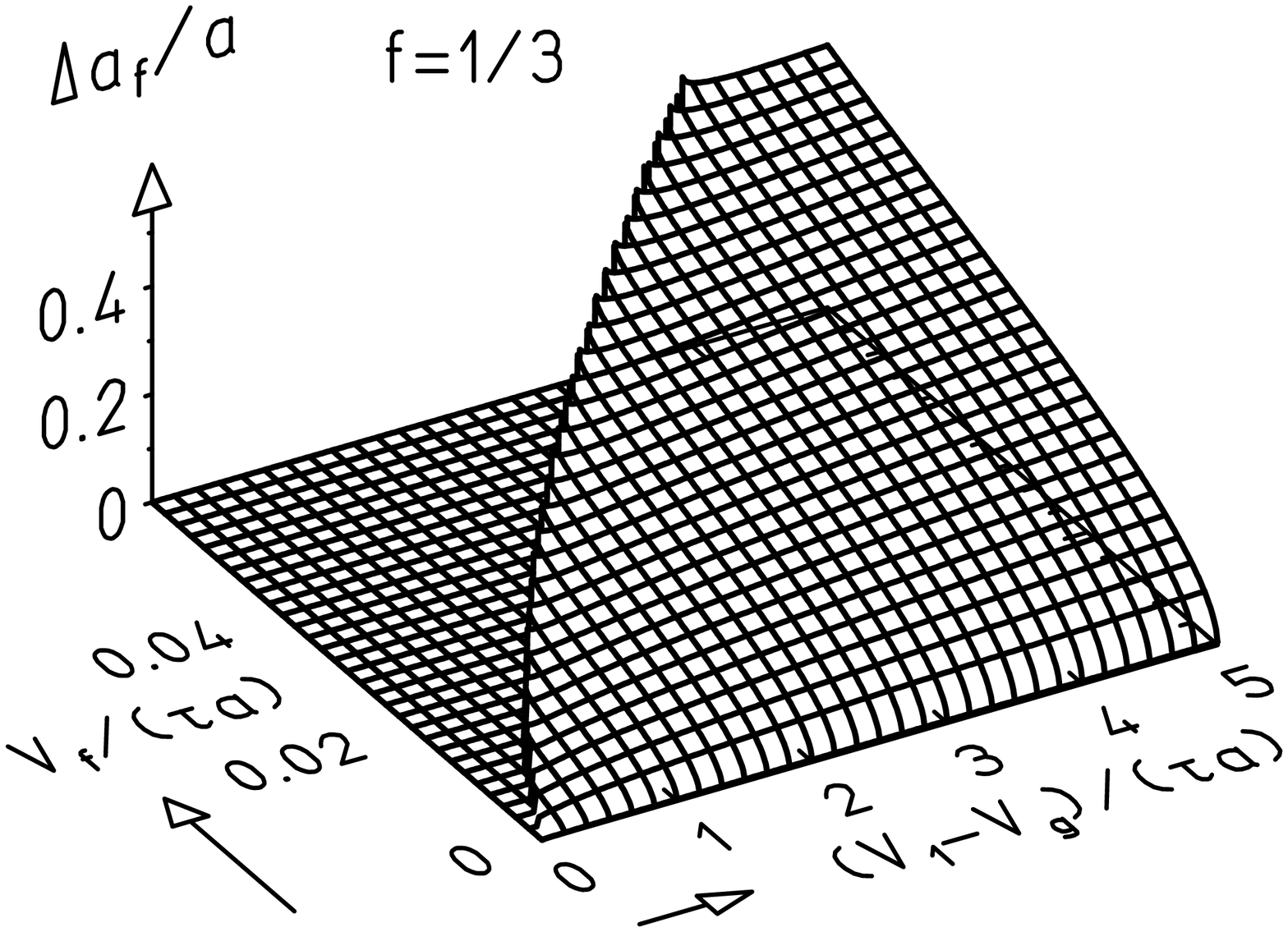}
\bigskip\par
\includegraphics[width=0.6\columnwidth]{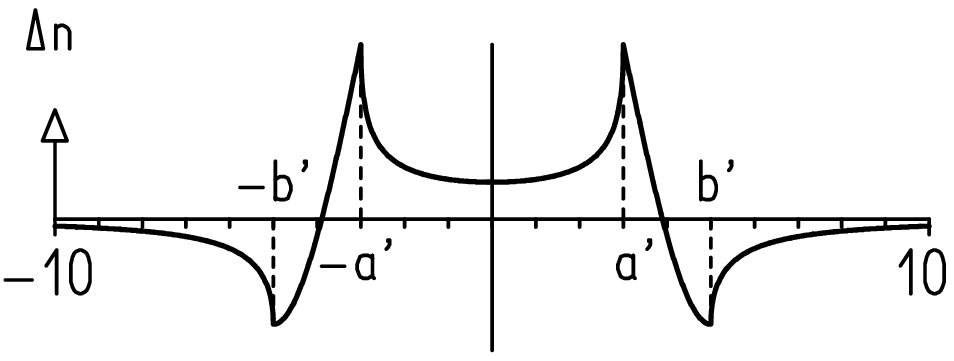}
\end{center}
\caption{Top: three-dimensional plot of the dependence of the
incompressible-stripe width $\Delta a_f$ for $f=1/3$,
formed at the edge of the 2DEGs in the presence of magnetic fields,
on gate voltage $V_1-V_\abbr{g}$ and charge gap $V_f$
for the fourth order of approximation of $n(x)$ around points $x_{f\pm}$.
For large values of $V_f$ no solution exists.
Bottom: Schematic illustration of fourth order for $\Delta n$.
($a'=3$, $b'=5$, coefficients of $\tau(x)$:
$\tau_2^{(2)}=1.25$, $\tau_4^{(2)}=-0.025$.)
}
\label{fig:8_sec_order}
\label{fig:Delta-n-2nd}
\end{figure}

Here we require in the approximation also the second derivative of
$n(x)$ at $x_f$ to coincide,
\begin{eqnarray}
\nonumber
  n(x)&\approx& n_f-\frac{1}{8} (5x_f n_f'-x_f^2 n_f'')
   +\frac{1}{4}\left(3\frac{n_f'}{x_f}-n_f''\right)x^2 \\* 
   &&{}+\frac{1}{8}\left(\frac{n_f''}{x_f^2}-\frac{n_f'}{x_f^3}\right)x^4
  \quad,
  \label{eq:n(x)-2nd-ord}
\end{eqnarray}
where $n_f''\equiv\partial_x^2n(x)|_{x_f}$.
The equations for the stripe edge positions will be given by a set of
equations similar to (\ref{eq:incstr-eqs-1st}),
where the $\tau_i^{(2)}$ denote the coefficients of
$\tau(x)= \sum_{i=0}^{N}\tau_i^{(N)} x^{2i}$  (with $N=2$ here)
in Eq.~(\ref{diel_stripes_new}) according to Eq.~(\ref{eq:n(x)-2nd-ord}):
{\small
\bea
  \!\!\!\!\!\!&&\frac{b'^2-a'^2}{2}\left(\tau_0^{(2)}
   +\tau_2^{(2)}\frac{a'^2+3b'^2}{4}
   +\tau_4^{(2)}\frac{a'^4+2a'^2b'^2+5b'^4}{8}\right)
   \nonumber\\*
  \!\!\!\!\!\!&&{}=\frac{V_f-A^{(2)}(a',b')}{\bK'}b'
   \;,
\label{two_eqs:sec_order}
\eea}
where again the second equation follows by the exchange
$a' \leftrightarrow b'$
on the left-hand side of Eq.~(\ref{two_eqs:sec_order}).
Difference and sum of these equations are given in
Appendix \ref{appx:2nd-order-supp}.
Their numerical solution is shown in Figs.~\ref{fig:8_sec_order} (top)
and \ref{fig:1st+2nd-order-cuts}.
A solution exists for not too large values of $V_f$ because of
$n_f'>0$ and $n_f''<0$, hence $\tau_2^{(2)}>0$ and $\tau_4^{(2)}<0$.
This is because the fourth order of approximation of $n(x)$ has an
upper bound (as the original profile Eq.~(\ref{profile_eq}), but
presumably at slightly smaller values) in contrast to the second
order (\ref{eq:n(x)-1st-ord}), which is unbounded at large values.
Therefore, a solution for Eq.~(\ref{eq:incstr-eqs-1st}) can be found for
any positive $V_f$, whereas for Eq.~(\ref{two_eqs:sec_order}) it cannot.

In fourth 
order the additional charge $\Delta n$ is given by
{\footnotesize
\bea
  \!\!\!\!\!\!\!\!&&\Delta n = \frac{\tau_2^{(2)}}{\tau_0}
   \left(\frac{a'^2+b'^2}{2}-x^2\right)
   +\frac{\tau_4^{(2)}}{\tau_0}
   \left(\frac{3a'^4+2a'^2b'^2+3b'^4}{8}-x^4\right)
  \nonumber \\*
  \!\!\!\!\!\!\!\!&&{}+ \frac{1}{\tau_0}\left(\tau_2^{(2)}+
  \tau_4^{(2)}\left(x^2+\frac{a'^2+b'^2}{2}\right)\right)
  \left\{ \begin{array}{lcl}
    \left|D'(x)\right| &:& |x|>b' \\[0.5ex]
    -\left|D'(x)\right| &:& |x|<a'
  \end{array} \right\}
  \nonumber
\eea}%
with $\tau_0=2\pi e/\epsilon$, a plot of which on a magnified scale
 along the edge is shown in Fig.\ \ref{fig:Delta-n-2nd} (bottom).

\begin{figure}
\begin{center}
\includegraphics[width=0.70\columnwidth]{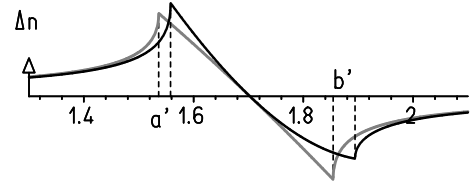}
\end{center}
\caption{Schematic illustration of second
(gray) and fourth
order (black) approximation for $\Delta n$
around the right IS. $a'$ and $b'$ are numerically determined
for magnified values of $n(x)$, $V_f/\tau'$.  The fourth order of
approximation leads to slightly wider ISs than the second-order one.  }
\label{fig:orders-schematic}
\end{figure}

\begin{figure}
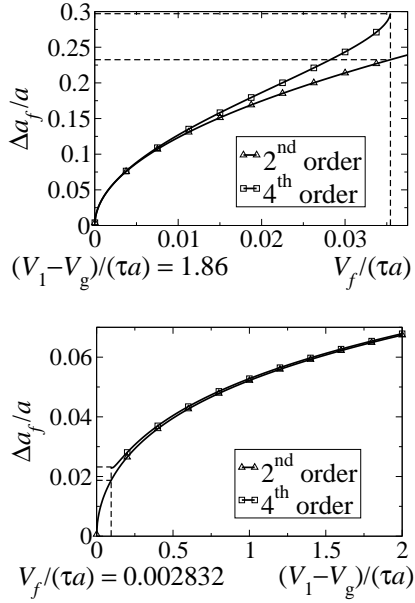

\begin{center}
\includegraphics*[width=0.62\columnwidth]{Fig11a.eps}
\bigskip\par
\includegraphics*[width=0.62\columnwidth]{Fig11b.eps}
\end{center}
\caption{Dependence of the width $\Delta a_f$ of the incompressible
stripe, of filling fraction $f=1/3$, on the magnitude of the charge
gap $V_f$ (left),  and on $V_1-V_\abbr{g}$ (right)
for second (triangles) and fourth
order (squares) of approximation of the
charge density around points $x_{f\pm}$.
Notice the square-root dependence of $\Delta a_f$ for small $V_f$.
At large gate voltages, $\Delta a_f$ also shows a square-root
dependence on $V_1-V_\abbr{g}$.
A solution in fourth
order for large values of $V_f$ compared to
$V_1-V_{\rm g}$ does not exist, see also Fig.\ \protect\ref{fig:8_sec_order}.}
\label{fig:1st+2nd-order-cuts}
\end{figure}

In the following we give comparisons of sizes of incompressible
and compressible-stripe widths between calculations of the second
and fourth
order. These two forms of approximation for $\Delta n$ are plotted in
Fig.\ \ref{fig:orders-schematic}.

The fourth-order calculations depart from the second-order ones
only 
signifcantly when $V_f/(V_1-V_{\abbr g}) \approx 0.01$, 
Fig.~\ref{fig:1st+2nd-order-cuts}. The experimental values of the
ratio of these voltages are $V_f/(V_1-V_{\abbr g}) \approx 0.0015$,
and so the fourth order will not give appreciable corrections
as shown in the table below. 
\begin{center}
\renewcommand{\arraystretch}{1.25}
\begin{tabular}{l|l|l}  \hline\hline
                                  & $a = 2500 \,{\rm \AA}$  &  $a = 5000\, {\rm \AA}$   \\[0mm] 
                                  & $V_\abbr{g} = - 0.674 \,{\rm V}$     & $V_\abbr{g} = - 0.674 \,{\rm V}$   \\[0mm] \hline
   $\nu=1$                        & $b = 4964 \,{\rm \AA}$            & $b = 7649\, {\rm \AA}$                \\[0mm] \hline
   $f=1/3$                        & $x_{1/3\pm} = \pm 2.076 \,a$       & $x_{1/3\pm} = \pm 1.544 \,a$          \\ [0mm]
$2^{\rm nd}\; {\rm order}$        & $\Delta a_{1/3} = 164.2\,{\rm \AA}$ & $\Delta a_{1/3} = 179.4\,{\rm \AA}$   \\[0mm] 
$4^{\rm th}\; {\rm order}$        & $\Delta a_{1/3} = 165.4\,{\rm \AA}$ & $\Delta a_{1/3} = 180.3\,{\rm \AA}$   \\[0mm] \hline
$f=2/3$                           & $x_{2/3\pm} = \pm 2.509 \, a$     & $x_{2/3\pm} = \pm 1.847\, a$         \\[0mm]
$2^{\rm nd}\; {\rm order}$        & $\Delta a_{2/3} = 338.0\,{\rm \AA}$ & $\Delta a_{2/3} = 375.9 \,{\rm \AA}$  \\[0mm] 
$4^{\rm th}\; {\rm order}$        & $\Delta a_{2/3} = 338.9\,{\rm \AA}$ & $\Delta a_{2/3} = 376.8\,{\rm \AA}$   \\[0mm] \hline\hline
\end{tabular}
\end{center}
The fourth-order approximation, however, is expected to give noticable contributions in the case of edges of higher Landau 
level fillings when $V_f$ is one order bigger than the case of the integer filling discussed here.

If the voltages $V_f$ were on the order of Landau level splitting (about 10 times larger than the current $V_f$), 
which is the case 
of edges of higher integer filling fractions, then the IS widths in
$2^{\rm nd}$ order of approximation for $f=1/3$ and $f=2/3$ had a size of
519.6\,\AA\ and 1071\,\AA, respectively. There the $4^{\rm th}$ order would
lead to sizable effects, leading to 582.6\,\AA\ and 1106\,\AA, respectively,
cf.\ also Fig.\ \ref{fig:1st+2nd-order-cuts}. 

Because $b$ grows almost linearly with $a$, the changes of the widths of the ISs are rather moderate as $a$ is varied.

In experiments \cite{pellegrini} the split gates are slightly out of the 2DEG plane ($\approx 10\,l_B$ off plane in
$z$ direction). 
As a consequence the gates provide a smoother confining potential than the coplanar case examined here.
This should be expected to lead to smoother edges and wider ISs. 

\section{Experimental Implications}

In the following we discuss implications arising from, respectively, the composite multichannel edge structure surrounding 
the split-gate constriction at high $|V_{\abbr g}|$ in QH systems at integer bulk fillings, 
and the inhomogeneously interacting non-chiral LL system in which they turn to at low $|V_{\abbr g}|$, in 
transport experiments.
Their observation would offer a further
proof of the consistency of the model that we propose to explain the nature of the 
metal--insulator transition observed in Ref.\ \onlinecite{pellegrini}.

At low $|V_{\abbr g}|$, measurements performed over a broad range of filling fractions  $1/3 \le \nu \le 1$ 
(Refs.\ \onlinecite{pellegrini,pellegrini_0,heiblum})
consistently observe a QP backscattering suppression across a constriction.  At these voltages the separation of 
counter-propagating edges under the gate is small and consequences of interedge interactions correspondingly are expected to be
most significant. 
The mechanism that we proposed therefore in Ref.\ \onlinecite{papa2}
and applicable for experiments \cite{pellegrini} where $\nu=1$,
attributes backscattering suppression to interactions near the point contact. This 
can be most easily understood by regarding the edges on opposite sides of the line junction as counterpropagating 
channels of a non-chiral LL defined on the constriction. These 1D systems form an inhomogeneous LL with parameter $K<1$ in 
the junction region (enforcing relevancy of left--right electron tunneling or suppression of top--bottom QP tunneling) 
and $K=1$ in the remaining portions of the edges.  These systems have properties that can be observed 
in experimental studies as we elaborate in Sect.\ \ref{sect:Low_Gate_Voltage}, below.

At high $|V_{\abbr g}|$ it is found in Sect.\ \ref{section:nature} that interactions across the line junction 
cannot lead to the opposite phenomena of enhancement of QP backscattering.
We explain below that in the reconstructed edge at high $|V_{\abbr g}|$
relevant fractionally charged QP tunneling processes arise and an
insulating behavior is to be expected. Our proposal for the explanation
of experimental observations at high $|V_{\abbr g}|$ is captured by the
schematic illustration represented in Fig.~\ref{charge_striped_magn}.

We comment in the following also on the features one should expect from noise experiments on these systems at high and low $|V_{\abbr g}|$.
 
\subsection{High Gate  Voltage} 

\begin{figure}
\unitlength=1mm
\begin{center}
\includegraphics[width=0.65\columnwidth]{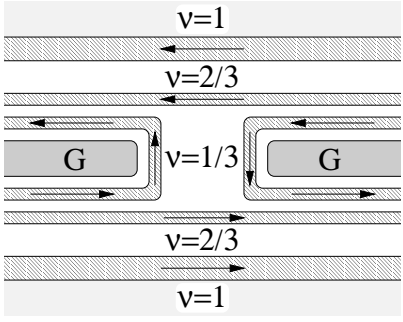}
\end{center}
\caption{
Schematic illustration of high $\left|V_\abbr{g}\right|$ splitting
of a $\nu=1$ edge into channels of fractional filling in QH
systems with constriction, similar to those employed in experiments of
Ref.~\onlinecite{pellegrini}.  }
\label{edge_config}
\label{charge_striped_magn}
\end{figure}

Detailed calculations in Sect.\ \ref{sect:edg-struct-B-pres} found that
for the geometry of the experimental system and the values
of $|V_{\abbr g}|$ that are used, 
the $1/3$ and $2/3$ ISs are of widths $2\,l_B$ and $4\,l_B$, respectively,      
separating CSs of sizes $2\,l_B$ and $10\,l_B$, respectively.                 
These ISs are wide when compared with the width of electron wave functions ($l_B$ for the noninteracting case but less than $l_B$ for the interacting one) 
in lowest Landau 
level states. Therefore wave functions for electrons in guiding centers 
localized in front of each other on opposite sides of the ISs, have an
overlap of the order of $\eun^{-\Delta a_{1/3}^2/l_B^2}$, 
where $\Delta a_{1/3}$ is the width of the narrower $1/3$ IS,
leading to small inter-compressible-channel tunneling amplitudes. 
Moreover, wide ISs enforce an effective statistics for the
fluctuating CS channels similar to the one characterizing
the $\nu=1/3$ edges.\cite{vignale}
In this paper, therefore, we propose the model depicted schematically in Fig.~\ref{charge_striped_magn} as the one that develops in
the experiments at high $|V_{\abbr g}|$ and that explains the insulating behavior across the constriction.
In the absence of interchannel Coulomb interactions, the
LL formed between $\nu=1/3$ channels at the line
junction anticrossing the Hall bar is equivalent to a
LL along the junction at filling $1/\nu$,\cite{papa2,renn}
a reflection of quasiparticle--electron duality in tunneling processes.
The flow of  $G$ from high to low $V$ can be described as follows: 
Starting with an open constriction (Fig.~\ref{edge_config}), slowly turning off $V$, the system flows from relevant
quasiparticle backscattering, suppressing the
source-to-drain conductance according to
$G- \nu e^2/h \sim - V^{2\nu-2}$,
to a closed constriction with irrelevant electron tunneling with $G\sim V^{2/\nu-2}$,
unable to open the constriction at small $V$.
In the absence of interactions the system is insulating at low bias
with $I$--$V$ characteristics of a QH system of bulk $\nu=1/3$.
The $I$--$V$ evolution being reminiscent of a bulk filling $1/3$
was indeed one of the puzzles of the experiments.\cite{pellegrini}

The chiral channels constituting the integer edge are in close proximity, nevertheless, and interactions cannot be neglected.
We show now that Coulomb interactions between channels do not influence 
the tunneling exponents.
In the following we adopt a model of coupled chiral channels, each
described by a single chiral LL, an assumption best fulfilled by
the outermost sharper channels.
The excitation spectrum of this model is composed of two slow neutral modes and a single faster
charged mode, each extended on all channels.
The action that accounts for the interchannel Coulomb interactions of same-chirality 
branches is
\be
 S = \sum_{i,j=1}^3 \int  \rd x \, \rd \tau
  \left[ \frac{2\imu\delta_{ij}}{\nu}
  \,\p_\tau \phi_i \, \p_x \phi_j + V_{ij} \,
  \p_x \phi_i \, \p_x \phi_j \right]
  \, ,
\ee
where  $\phi_i$, $i=1,2,3$, are bosonic fields associated with each CS on one side of the gates.
$\p_x \phi_i =-\sqrt{\pi}\rho_i(x)$ is the charge density of channel $i$ and quantization is assumed
based on the commutation relations
$[\rho_i(x),\rho_j(x')]=-({\imu\nu}/{2\pi}) \delta_{ij} \p_x\delta(x-x')$.
We can assume in general the intrachannel interactions $V_{ii}$
for $i=1,2,3$ to
be different, and the interchannel ones to fulfill $V_{ij}=V_{ji} \neq V_{ii}$.
The diagonalized action can be written  as
\bea
S= \sum_{i=1}^3 \int \rd x \, \rd \tau
\left[\frac{2\imu}{\nu} (\p_\tau {\varphi}_i)(\p_x {\varphi}_i)
 + \lambda_{i} (\p_x {\varphi}_i)^2  \right]  \; ,
\eea
where ${\varphi}_i=M_{ij}\phi_j$ and $\lambda_i$ are eigenvalues of $V$.
Since all channels are of the same chirality, the interaction-dependent $M$ matrix is
unitary, $(M^\dagger M)_{ij}=\delta_{ij}$, as opposed to the case corresponding to the presence of
opposite-chirality channels in which case $M^\dagger J  M = J$. 
The matrix elements of $J$ are $(J)_{ij}=\delta_{ij}\,{\rm sign}(j)$ and $J$ contains at least 
one element of opposite sign compared to the rest, corresponding to the opposite-chirality channels present. 
This will result in $(M^\dagger M)_{ij} \neq \delta_{ij}$, thus affecting scaling exponents of tunneling processes.
The correlation functions between electrons
of branch $i$,
$R_i^{\rm el.} \sim \allowbreak \eun^{\imu \sqrt{4\pi}\phi_i/\nu}$,
have the form 
\bea
\bigl< {R_i^{\rm el.}}^\dagger (\tau) R_i^{\rm el.}(\tau')\bigr>
\sim \allowbreak \prod_{j=1}^{3}(\lambda_i (\tau -\tau'))^{-2 (M^\dagger M)_{ij}/\nu} \; ,
\eea
and the scaling dimensions are $d_i=\sum_j (M^\dagger M)_{ij}/\nu=1/\nu$, irrelevant for $\nu=1/3$
and unaffected by the interactions.

Another point to consider is the question of stability of the channel structure under interchannel 
tunneling processes mediated by impurities along the edge.  For repulsions $V_{ij}=V$  ($\le V_{ii}=U$),
the action is diagonalized in terms of two neutral and one charged field.
At low energies the neutral fields tend to antilock charge densities in the consecutive channels leading to 
an oscillatory charge density profile across the edge ($x$ direction), in analogy to semi-classical results obtained in Ref.\ \onlinecite{vignale2}.
Interchannel tunneling processes lead to the chiral sine-Gordon model
$\cos({\sqrt{8\pi}}\varphi_n/\nu)$, $n=1,2,$ for neutral fields. (In fact this does not depend  on $V_{ij}$.)
The ${\it cos}$ operator is relevant only for $\beta^2 \le 16\pi$,\cite{sondhi}
a condition not realized in these systems since $\beta^2=8\pi/\nu^2$
and $\nu =1/3$.

\begin{figure}
\unitlength=1mm
\begin{center}
\includegraphics[width=0.75\columnwidth]{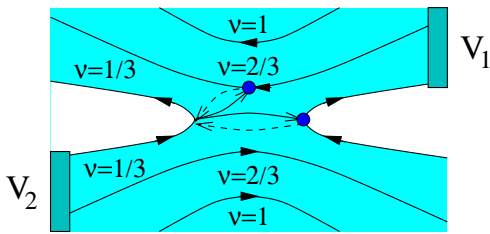}
\end{center}
\caption{ 
Schematic illustration at high $\left|V_\abbr{g}\right|$ and large bias of the tunneling processes;
cross-current correlations, that can in principle be measured in finite-frequency current-noise 
experiments. Such processes are influenced by the charge and statistics of the quasiparticles that tunnel and can be used 
to detect their statistics (Ref.~\onlinecite{kim2005}). 
}
\label{qp_statistics}
\end{figure}

The multichannel edges that appear in the insulating state at high $|V_{\abbr g}|$ and high $V$ 
offer a realization of models suggested in recent papers by Kim {\it et al.},\cite{kim2005} where 
finite-frequency current-noise experiments measuring cross-current correlations could observe 
the fractional statistics of QPs participating in the tunneling processes.
The phase of the relative wave function of two QPs picked up upon their exchange can be observed in the proposed noise experiments if their  
exchange is mediated by a third channel. The systems studied in this paper provide such a channel naturally.
Care, nevertheless, must be taken to localize the tunneling processes to a small region
at, for instance, near the constriction  as
depicted schematically in Fig.~\ref{qp_statistics}, 
by placing additional gates to
increase the distance of the edges along the rest of their lengths.

As $\left|V_\abbr{g}\right|$ is lowered we expect the $1/3$ IS channel to eventually close
with the remaining $2/3$ IS to extend under the gate.
At large bias, the same argument as in {\it high\/} $\left|V_\abbr{g}\right|$ for the conductance applies,
with $G - \nu e^2/h \sim - V^{2\nu -2}$ and $\nu=2/3$.
At low $V$, however, Coulomb interactions between oppo\-site-chirality channels as discussed
in Ref.\ \onlinecite{papa2} are expected to be strong leading to perfect transmission through
the constriction and exhibiting the symmetry $\nu=1/3 \leftrightarrow \nu=2/3$ between
the low and high $|V_{\abbr g}|$ observed in the experiments of Ref.~\onlinecite{pellegrini}.

One point that needs further comment here is the necessity of comparing this multichannel edge description with 
the one of the Hall systems of 2/3 bulk filling which is known to contain counterpropagating modes 
\cite{wen,macdonald} when electron correlations are taken into account.
Based also on recent density functional theory (DFT) calculations on the edge structure in QH systems of integer filling, 
around an antidot,\cite{ihnatsenka} 
such modes do not seem to arise at the edges of 2/3 ISs formed at the integer edge. Therefore our 
description, we believe, holds even if electron correlations are taken into account.

\subsection{Low Gate Voltage}  

\label{sect:Low_Gate_Voltage}

The experimental findings of suppression of backscattering between the top and bottom edges in QH systems with gate-created constrictions at small values of 
$|V_{\abbr g}|$, whether at bulk filling $\nu=1/3$ or at integer filling $\nu=1$, according to our model, is the consequence of 
strong Coulomb interedge repulsions across the junction region.  
Features that can be best exploited for a
consistency check of this proposal (of interest also in the broader context of observation of the LL model) 
are the ones that depend on the LL parameter $K$. The ones that we discuss here
are two: 
a $K$-dependent oscillatory behavior of $I$--$V$,\cite{dolcini} and the
formation of a particle--hole order parameter in the junction region. 

\begin{figure}
\unitlength=1mm
\begin{center}
\includegraphics[width=0.65\columnwidth]{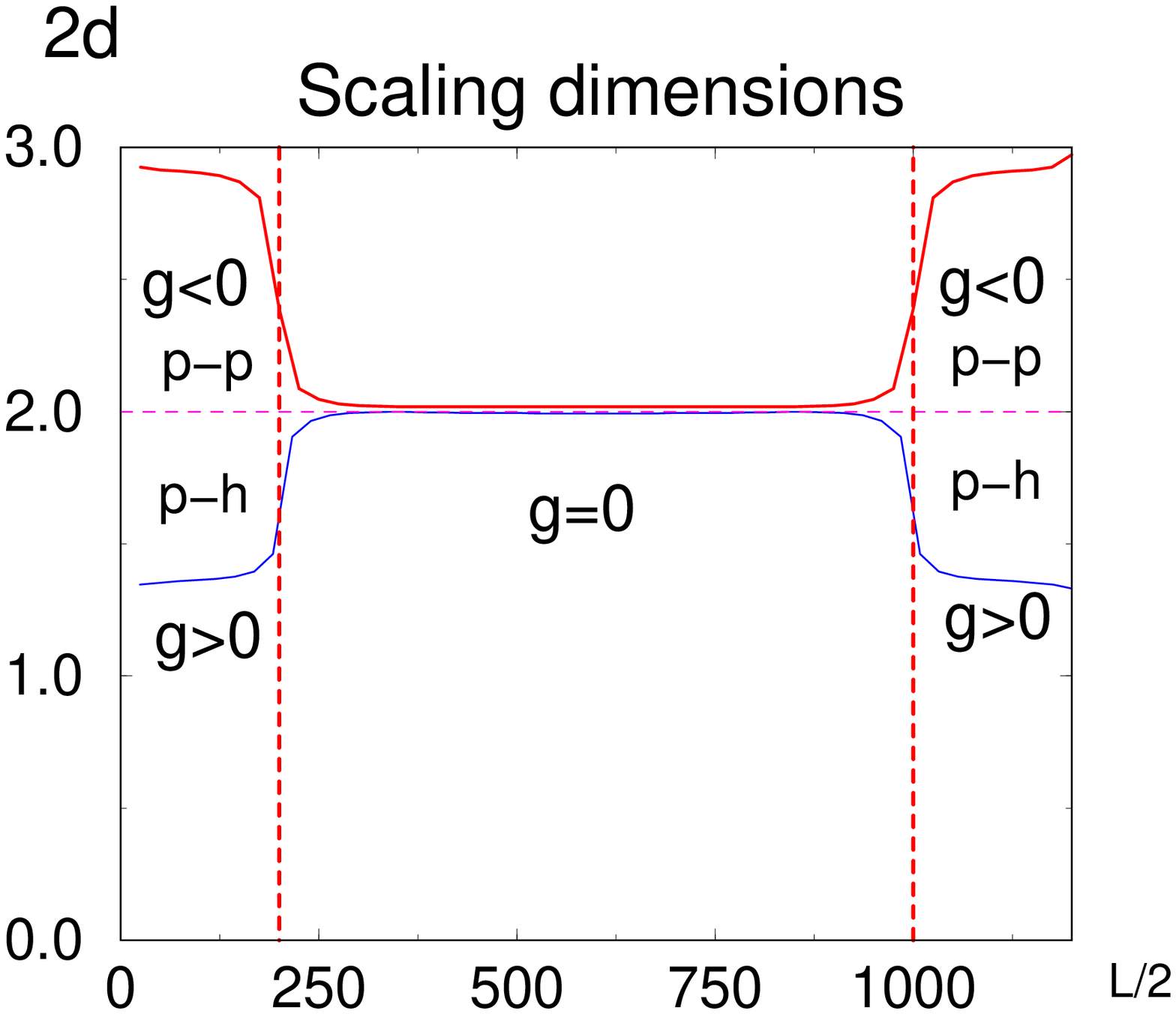}
\par\bigskip
\includegraphics[width=0.40\columnwidth]{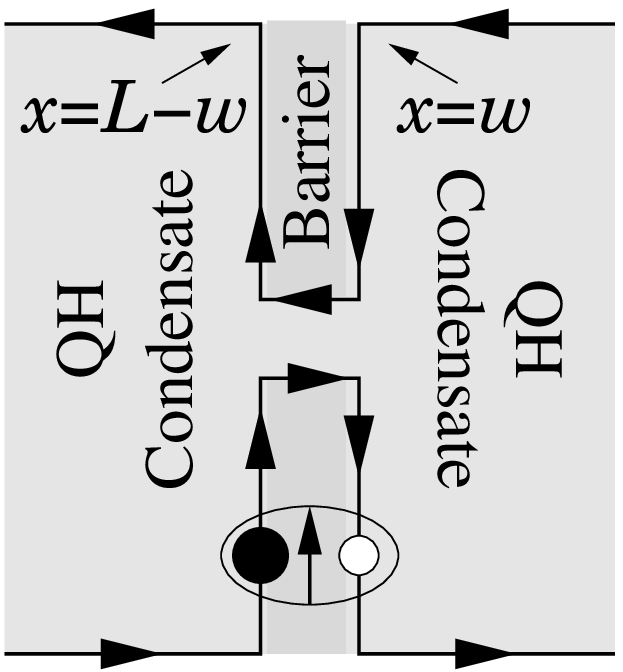}
\end{center}
\caption{ 
Scaling dimensions of the particle--hole order-parameter operator
of an inhomogeneous Luttinger liquid. The length of the system is
$L=2400\,l_{B}$.
The fluctuations are relevant only in and just outside the regions of repulsive interactions.
Interactions take place on both ends of the edge loop in a length of $w=200\, l_{B}$ each.
}
\label{fig:sc-dim-reg}
\end{figure}

The interedge Coulomb interaction in systems of concern here is much
stronger in the junction region than on other portions of the edge.
Therefore, in these systems the interactions define a piecewise
Fermi velocity $v_{\rm F}$ along the length of the non-chiral LL,
formed by the counterpropagating edges on the opposite sides of the
junction, $v_{\rm F}=v_{\rm F}^0/K$ in the junction region and
$v_{\rm F}^0$ in the other portions of the edge.
Discontinuities in the value of $v_{\rm F}$, on the other hand, lead
to physically observable consequences in transport experiments. 
Namely, the requirement of continuity of the current at all points
along the LL leads to the phenomena of partial Andreev reflections 
\cite{safi} of edge waves at the interaction inhomogeneity
boundaries ($x=\pm w$ or $x=L/2\pm w$, where $L$ is the perimeter of the
QH edge and $w$ the length of each gate, Fig.~\ref{fig:sc-dim-reg}).
On the other hand, the finite length of the QH line junction 
introduces a time (or frequency $\omega_{\rm LJ} \sim v_{\rm F}/2w$) scale;
the time for the edge waves reflected from the impurity to move to the
junction boundary, where they incur further reflection, and come back again.
The superposition at the point contact of these multiply reflected waves and the incoming ones 
can be tuned to constructive or destructive interference by, either a
variation of 
strength of the Coulomb interactions
(thus $v_{\rm F}$ and $\omega_{\rm LJ}$) that can be tuned by
changing the separation distance between the left and right edges, 
or by a variation of the length of the gates, leading to oscillatory
behavior of the frequency $\omega \sim \omega_{\rm LJ}$ of the current
transmitted through the constriction.
The intensity and amplitude of these oscillations, depending on the
amplitude of the multiple reflections of the edge waves, varies as
a function of $K$, too. Oscillatory behavior in $I$--$V$ was observed
in fact in the experiments \cite{pellegrini} and \cite{pellegrini_0} 
although a detailed analysis of its frequency dependence on junction
length $\Lambda=2w$, interaction strength, 
and most importantly a distinction from oscillatory behavior
originating from other sources, 
unrelated with the presence of electron correlations in the systems
(like Fabry--P\'erot ones \cite{liang}, etc.), for purposes of interest
here remains to be done.

The other point of physical importance that arises due to electronic correlations in the 
junction region, is the formation of the  particle--hole order parameter there: 
\bea
  \hat{O}(x,t) = {R^{\rm el.}}^\dagger(x,t) {R^{\rm el.}}(-x,t) \quad .
  \label{p-h-operator}
\eea
The origin of these phenomena is the drag effect induced by a propagating charge density on one side of 
the barrier to the other side to which it is coupled by Coulomb interactions. 
Formally this appears in the splitting of the bosonic fields into left--right movers that physically extend 
on both edges.
The geometry assumed here is that of a single edge enclosing a 2DEG in a QH plateau of filling 
factor $\nu$, Fig.~\ref{fig:sc-dim-reg} (bottom). 
The Hamiltonian as usual is described in terms of charge density fluctuations ${\cal H}(x,x')=\rho(x)U(x,x')\rho(x')$, 
where $x$, $x'$ are coordinates 
along the perimeter of the edge and $U(x,x')=\left.\delta^2 E/\delta \rho(x) \delta \rho(x')\right|_0$, 
is the interaction kernel. 
Interedge interactions at low $|V_{\abbr g}|$ are of Coulomb type, 
$V_{\abbr d}(x)=e^2/(\epsilon [x^2+ \alpha l_B^2]^{1/2})$
and are concentrated only on the  region of the gates, $x\in (L-w,w)$ and $x\in (L/2 - w,L/2 + w)$, 
where $L$ is the edge perimeter length.
$\alpha$ is a dimensionless constant of order one and serves as a short-distance cutoff parameter. 
Interactions on the same side of the barrier, $V_{\rm s}(x)$, are also of Coulomb type.
Here of importance is   
the exponent of the expected algebraic decay with time of 
$\left<\hat{O}^\dagger_\abbr{}(x,\tau) \hat{O}_\abbr{}(x,0)\right>$, which we calculate numerically. 
To this end the Hamiltonian is diagonalized in momentum space with matrices 
$M^\dagger$, $M$, such that $M^\dagger J M=J$, where matrix $J$ has elements different from zero only along 
its main diagonal, unity in absolute value but half of them being negative. 
The size of the matrices chosen in 
our numerics is of the order of hundreds.
Using the identity of Ref.\ \onlinecite{identity} for the scaling
dimensions $d$ of the particle--hole order parameter, we obtain
\bea
  d = \hspace{-4mm}&&-\frac{1}{2}\frac{4\pi}{\nu^2}\Bigl\{\frac{1}{L}\sum_{k,m=-\infty}^{\infty} 
  \frac{1}{2\sqrt{|q_kq_m|}}8 \sin(q_k x)\sin(q_m x) 
  \Bigr.
\nonumber   \\[0mm]
  &&\Bigl. \frac{}{}
   M(q_k,q_l)\bigl[N(q_l;0) - N(q_l;\tau-\tau')\bigr] M^\dagger(q_l,q_m) \Bigr\}
  ,\qquad
\eea
where $q_k=2\pi k /L$ with $k$ integers $k=0,\pm1 , \pm 2 \ldots$ 
(periodic boundary conditions were assumed) and  
$N(q;\tau-\tau') = \exp\{-\lambda_q |\tau-\tau'|\}/2$ and $\lambda_q$ are energy eigenvalues 
(eigenvalues of $JH$). Numerical results for the scaling dimensions of the particle--hole order parameter are shown in 
Fig.~\ref{fig:sc-dim-reg} (top).

In cases of closed constrictions, calculations can be done similarly. 
It is expected\cite{pryadko} that left--right electron tunneling in the
closed constriction limit is dual to the top--bottom interedge tunneling 
of the open constriction limit.
In this case correlations of constituent operators have to be evaluated on each separate side of the gate.
In case the interactions are over an extended region, the $\lambda_\pm(q)$, where $\pm$ refers to the left and right
Hall system on the left and right of the gates, each of filling $\nu_1$ and $\nu_2$, are given by
\bea
\lambda_\pm(q) = &&
\left(\left[{ V}^2_\abbr{s}(q) \Bigl(\frac{\nu_1 + \nu_2}{2}\Bigr)^2 - \nu_1\nu_2 {V}^2_\abbr{d}(q) \right]^{1/2}
\right.
\nonumber \\[1mm]
&& \left.
\pm \frac{{ V}_\abbr{s}(q)}{2}\left[\nu_1 - \nu_2 \right] \right)|q|.
\eea
If $\nu_1=\nu_2$ then $\lambda_\pm(q)\equiv \lambda(q)$, with  
$\lambda(q) = \nu [ { V}^2_\abbr{s}(q) - {V}^2_\abbr{d}(q)  ]^{1/2} |q|$.
In general, however, calculations for eigenvalues have to be done numerically. 

The particle--hole order parameter is a relevant operator in the interaction region and 
it persists to be so even in a small interval outside this region, see Fig.~\ref{fig:sc-dim-reg}. The relevance of this operator 
depends on the strength of the interactions. 
Experiments probing the distribution of charge of waves incident on the
line junction like the ones discussed in Ref.~\onlinecite{lee2005}, which can
lead to the observation of the particle--hole order in and around the gate
region, can possibly measure $K$ as well as its $V_{\abbr g}$ dependence.

\section{Conclusions}

In the current paper we have explored the reasons for the occurrence of the insulating behavior at 
high $|V_{\abbr g}|$
at a constriction created by electrostatic gates in a 2DEG in which the integer QHE has been established.
First we examined the nature of interactions between edges in the constriction region. 
More precisely, we answered the question of how image charges appearing on the gates influence the interedge interactions. We find that at low $|V_{\abbr g}|$
(small 2DEG--gate distances) the interedge interactions are well screened. At large values of 
$|V_{\abbr g}|$ a sign reversal of interactions does not occur;
interactions remain repulsive in nature at any voltage of 2DEG subsystems.  
Here the analytic electrostatic calculations have been done exactly with no approximations.
We then studied the structure of the gate-induced  edge in the presence of magnetic fields.  
Smooth edges have an instability toward formation of compressible and incompressible stripes, around points of 
fractional (or integer at higher) fillings. It turns out that the experimental geometry and values 
of gate voltages used in Ref.~\onlinecite{pellegrini} lead to formation of wide ISs   
whose presence as shown in Fig.~\ref{edge_config},
can lead to enhanced top--bottom fractional-charge backscattering
or, equivalently, to a suppression of left--right electron tunneling
through the constriction. 
Moreover, the $G$--$V$ characteristics of constriction tunneling are
expected to be similar to the one of bulk fillings $1/3$, as observed
in the experiments.\cite{pellegrini}
The consistency of models proposed here 
can be verified, at high $|V_{\abbr g}|$, in addition to noise experiments commented on in the last section, also
by equilibration experiments similar to those of Alphenaar
{\it et al.}\cite{alphenaar} One can inject current on one channel of the integer bulk edge by the use of 
additional metallic gates that separates only the outermost channel, for instance, bringing it in contact 
with a current source. One can measure its redistribution in time as well as its splitting as it passes 
through the point-contact region.
At low $|V_{\abbr g}|$, properties of the emerging inhomogeneous LL (discontinuity
in value of Fermi velocity) with its 
tunability of interaction strength by the voltage of the gate as well as the finite size of the junction region
can be exploited in consistency verification experiments.
The $K$-dependent oscillatory behavior of $I$--$V$, and the 
particle--hole order parameters were proposed as signatures to be observed in the experiments.

\section*{ACKNOWLEDGEMENTS}

Some of the results contained here were briefly discussed in a previous publication \cite{Papa_Stroh_PRL2006}.
We are grateful to J. Betouras,  M. Grayson, A. MacDonald, 
A. de Martino, V. Pellegrini,  M. Polini, S. Roddaro and A. Tsvelik for illuminating conversations and to Eun-Ah Kim for comments on the manuscript.  
E.P. would like to gratefully acknowledge the hospitality of the Theory Institute 
at Brookhaven National Laboratory where part of this work was completed.  
This work was supported by NSF Grant No.\ DMR-0412956.

\appendix{}

\section{Potential configuration of charged dielectric stripes}

\label{App_A}

\subsection{General form of charge distributions}

A 2D potential can be represented as the imaginary part of an analytic
function $F(\zeta)$ with $\zeta=x+\imu z$. For the derivative
${\rd F}/{\rd \zeta}={\p_z \IM F}+\imu{\p_x \IM F}$ 
holds and therefore the components of the electric field are
\be
  E_x
  =-\IM\left[\frac{\rd F}{\rd \zeta}\right]
  \quad,\quad
  E_z
  =-\RE\left[\frac{\rd F}{\rd \zeta}\right]
  \label{eq:el-comp-F}
  \quad.
\ee
The boundary conditions of Sect.\ \ref{sect:sym-case} are applied
to $\IM F \equiv \phi^{\rm str.}(x,z)$. In the depleted region there is
\bea
  \frac{\p \IM F}{\p z} = \RE\left[\frac{\rd F}{\rd \zeta}\right] 
  =\tau(x) \quad , \quad a < |x| < b \quad .
  \label{eq:sym-ins}
\eea
Since the function $\phi^{\rm str.}(x,z)$ is constant along the $x$ axis
outside the above intervals (at $z=0$), 
$E_x(x,0)=0$, i.e.,
\be
  \frac{\p \IM F}{\p x}= {\rm Im} \left[\frac{\rd F}{\rd \zeta}\right] = 0
  \;\; , \;\;
  |x| < a \;\;\mbox{or} \;\; |x| > b\;\;.
  \label{eq:sym-outs}
\ee

Now another function, $D(\zeta)$, is introduced whose imaginary part is
known everywhere along the $x$ axis. From (\ref{eq:sym-ins}),
(\ref{eq:sym-outs}) one knows
$\RE[\rd F/\rd \zeta]$ in part of the real $x$ axis, $a<|x|<b$, and
also $\IM[\rd F/\rd \zeta]$ in the remaining part of the
$x$ axis. Therefore, if one multiplies this function with another, known
one, which is purely imaginary in the intervals $a<|x|<b$ and real
otherwise, then a new function $f(\zeta)$ is built whose imaginary part
one knows everywhere along the $x$ axis.

Such an auxiliary function 
is $\imu [(x^2-a^2)(b^2-x^2)]^{1/2}$.
Therefore, the new function
whose imaginary part is known everywhere along the $x$ axis, is given by
\be
  f(\zeta) = \imu  D(\zeta) \, \frac{\rd F}{\rd \zeta}
  \;\;, \;\;
  D(\zeta)=[(\zeta^2-a^2)(b^2-\zeta^2)]^{1/2}\;\;.
  \label{f_function02}
\ee
We can generate the function in the whole complex plane by using the
Schwarz equality
\bea
  f(\zeta) = \frac{1}{\pi} \int_{-\infty}^{\infty} \rd x
  \frac{D(x)[\rd F/\rd \zeta](x)}
  {x-\zeta} \quad .
  \label{eq:Schwarz}
\eea
In our case $\rd F/\rd \zeta$ obeys
Eqs.~(\ref{eq:sym-ins}), (\ref{eq:sym-outs}).

\begin{figure}
\begin{center}
\includegraphics[width=0.6\columnwidth]{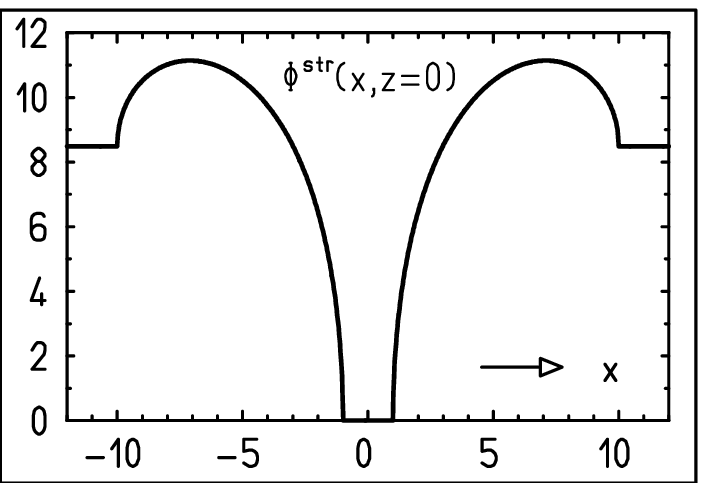}
\bigskip\par
\includegraphics[width=0.6\columnwidth]{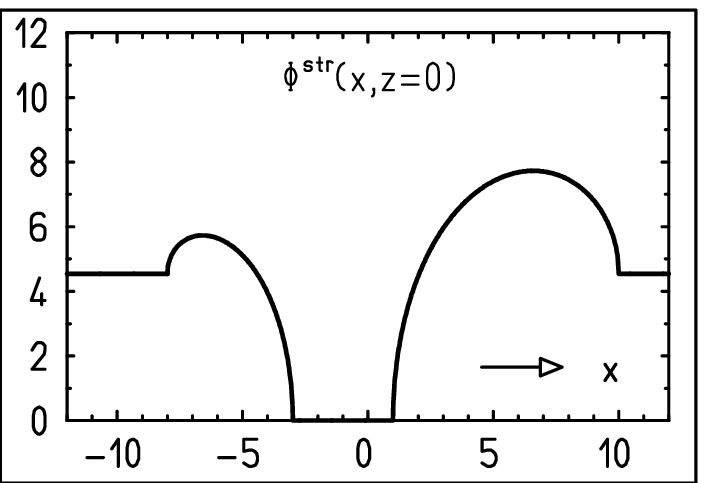}
\end{center}
\caption{Top: a plot of the potential $\phi^{\rm str.}(x,z=0)$ created
by a dielectric (\ref{stripes_pot}) for certain values of
$a=1\,l_B$, $b=10\, l_B$. Bottom: a plot of the potential
$\phi^{\rm str.}(x,z=0) = \IM F(\zeta)$, where $F(\zeta)$ is given
in (\ref{eq:asy-F-gen}). The selected values of $\bar b$, $\bar a$, $a$, $b$
are $\bar b=-8\,l_B$, $\bar a=-3\,l_B$, $a=1\,l_B$, $b=10\,l_B$.}
\label{fig15}
\end{figure}

The integral (\ref{eq:Schwarz}) just reproduces the numerator of its
integrand on the real axis including a factor $\imu$. For some analytic
$\tau(x)$ the result can be expressed as
\bea
  f(\zeta) =  \imu\tau(\zeta) D(\zeta) + {\cal F}(\zeta,a,b)
  \quad,
\label{eq:f=i(D+F)}
\eea
where ${\cal F}(\zeta,a,b)$ is also analytic and is to be found from
the boundary conditions. These are (a) $\IM[{\cal F}(x,a,b)]=0$ everywhere
on the real axis and (b) $f(\zeta)\to 0$ as $|\zeta|\to\infty$, which is
imposed by (\ref{eq:Schwarz}). For $\tau_n(x)=\tau_nx^n$ one obtains
\be
  {\cal F}_n(\zeta,a,b)=
   \tau_n\zeta^{n+2}\sum_{k=0}^{\left[\frac{n+2}{2}\right]}
   C_k^{-\frac{1}{2}}\!\left({\textstyle \frac{a^2+b^2}{2ab}}\right)
   \left(ab/\zeta^2\right)^k
  \,,
  \label{eq:F-4-mon-n}
\ee
where $C_k^{-\frac{1}{2}}(t)$ is a Gegenbauer polynomial,
cf.\ Eq.~(8.930) of Ref.~\onlinecite{gradshteyn}, and $[\cdot]$ represents the integer part.
For constant $\tau(x)$ the function $f(\zeta)$ results in
($C_0^{-\frac{1}{2}}(t)=1$, $C_1^{-\frac{1}{2}}(t)=-t$)
\[
  f(\zeta) =  \tau_0\left[\imu D(\zeta) + \zeta^2-\thalf(a^2+b^2)\right]
  \quad.
\]

The electric field is already gained from (\ref{f_function02})
via (\ref{eq:el-comp-F}),
\bea
  \frac{\rd F}{\rd \zeta} =
  \tau(\zeta)-\imu\frac{{\cal F}(\zeta,a,b)}{D(\zeta)} \quad,
  \label{eq:F-der-zeta}
\eea
which has the expected singular square-root behavior at every edge.
These are the only singularities caused by $D(\zeta)$, and since it
appears in the denominator, it should also not have any additional
zero in the complex plane. Together with the requirements imposed
for its construction, the function $D(\zeta)$ is (almost) fixed.

The function $F$ introduced is obtained by the integration of
(\ref{eq:F-der-zeta}) along a curve within the considered half-plane
from $a$ to $\zeta$.
These integration boundaries 
fix the boundary potentials outside the dielectric stripes to zero
in the gate region and $\phi^{\rm str.}(x,z=0)=A(a,b)$ in the 2DEG regions.
The potential constant $A(a,b)$ introduced in Eqs.~(\ref{met_components1}) and (\ref{outside_interv})
comes out to be:
\bea
\label{A(ab)}
A(a,b) = -\tau b \left[\bE\left(k'\right) - \frac{1}{2}\left(1 + \frac{a^2}{b^2}\right) \bK\left(k'\right) \right],
\eea
where $k=a/b$ and $k'=(1-a^2/b^2)^{1/2}$.

For the general case of Sect. \ref{section:nature}
the potential created by the dielectric stripes, after integration
of $f(\zeta)/D(\zeta)$,\cite{D_F,byrd} is obtained to be 
 \begin{eqnarray}
   \phi^{\rm str.}(\zeta) &=& \tau \, {\rm Im} \Bigl\{
 \frac{\imu}{4}\frac{(b-\bar a-a+\bar b)^2}{[{(b-\bar a)(a-\bar b)}]^{1/2}}
   F(\varphi,\tilde k) \Bigr. \\[0mm]
   &-&\imu [{(b-\bar a)(a-\bar b)}]^{1/2}\left(E(\varphi,\tilde k)
   -F(\varphi,\tilde k)\right)
   \nonumber \\[0mm] \nonumber
   &+&\Bigl.  \imu \left[{\frac{(b-\zeta)(\zeta-a)(\zeta-\bar b)}{\zeta-\bar a}}\right]^{1/2}
   + (\zeta-a)\Bigr\} \;.  
   \label{eq:asy-F-gen} 
 \eea
Here $\tilde k$ is the modulus of the elliptic functions,
$\tilde k=\left[{(b-a)(\bar a-\bar b)}/{(b-\bar a)(a-\bar b)}\right]^{{1}/{2}}$,
and $\varphi$ is given by
$\varphi=\arcsin\left[{{(b-\bar a)(\zeta-a)}/{(b-a)(\zeta-\bar a)}} \right]^{{1}/{2}}$.
[The singularities from the square-root term and from $\varphi$ in
(\ref{eq:asy-F-gen}) at $\zeta\to\bar a$ just cancel (Fig.~\ref{fig15}).]
It is worthwhile to notice that the modulus $\tilde k$ relates to $k=S_k^-/S_k^+=I(b)/I(a)$ of Sect. \ref{sect:noi-fld-2DEG} 
by $k=(1-\tilde k)/(1+\tilde k)$.

\subsection{Polynomial charge distributions}
\label{appx:pol-chrg-dist}
\label{appx:2nd-order-supp}

The charge density for ISs is given by the density profile $n(x)$
(\ref{eq:profile_eq}), see Sect.\ \ref{sect:edg-struct-B-pres}.
Approximating $\tau_0(n(x)-n_f)$, $\tau_0=2\pi e/\epsilon$, up to
terms containing the $N$th derivative of $n(x)$ at $x_f$ we arrive at
different polynomial approximations for intervals $a'<|x|<b'$, where we
denote by $\tau_i^{(N)}$ the coefficients of $x^i$ for a given order $N$.

For $N=1$ the (second-order) approximation is given by (\ref{eq:n(x)-1st-ord})
with $\tau_0^{(1)}=-\frac{1}{2}x_f\tau_0 n_f'$ and
$\tau_2^{(1)}=\frac{1}{2}\tau_0 n_f'/x_f$. The function $F$ from
(\ref{eq:el-comp-F}) is obtained by the integration of
(\ref{eq:F-der-zeta}) with terms ${\cal F}$ from
(\ref{eq:F-4-mon-n}) and $D(x)$ replaced by
$D'(x)=[(x^2-a'^2)(b'^2-x^2)]^{1/2}$, resulting in
\bea
F^{(1)}(\zeta) &=& \tau_0^{(1)}b' \left\{\tilde{\zeta}-\bar{k}-
   \imu  \left[ \left(\bE'-E(\psi,\bar k')\right) \right.\right.
  \label{eq:F^1}
  \\*
  &&\!\left.\left. {}-\thalf  (1 + \bar{k}^2) 
   \left(\bK'-F(\psi,\bar k')\right)\right]
   \vphantom{\tilde{\zeta}} \right\} \nonumber \\
  &&{}+{\textstyle \frac{1}{3}}\tau_2^{(1)}b'^3\left\{{\textstyle }
   (\tilde{\zeta}^3-\bar{k}^3)+\imu\left[
   {\textstyle }\tilde{\zeta} D'(b'\tilde{\zeta})b'^{-2}
   \right.\right.\nonumber\\
  &&{}+{\textstyle \frac{1}{2}}(1+\bar{k}^2)
   \left(\bE'-E(\psi,\bar k')\right) \nonumber\\*
  &&\!\left.\left. {}+{\textstyle \frac{1}{8}}(3\bar{k}'^4+8\bar{k}^2)
   \left(\bK'-F(\psi,\bar k') \right) \vphantom{\tilde{\zeta}}
   \right]\right\} \nonumber \; .
\eea
Here $\tilde{\zeta}=\zeta/b'$, $\bar k=a'/b'$
and $\psi, \bar k'$ are given by the same
equations as for (\ref{stripes_pot}), 
with $a$ and $b$ substituted by $a'$ and $b'$.
This leads to the potential difference $A^{(1)}$ (\ref{eq:A_1}).
The approximation for $N=2$ (fourth-order) is given in (\ref{eq:n(x)-2nd-ord}).
One can partially reuse the results for $N=1$ using different
coefficients $\tau_i$. For $F$ this leads to
\bea
  \lefteqn{F^{(2)}(\zeta) = F^{(1)}(\zeta)
  \Bigl\{\tau_0^{(1)}\to\tau_0^{(2)},
  \tau_2^{(1)}\to\tau_2^{(2)}\Bigr\}}\nonumber\\[0mm]\nonumber
  &&\!\!{}+ {\textstyle \frac{1}{5}} \tau_4^{(2)}b'^5\Bigl\{
   (\tilde{\zeta}^5-\bar{k}^5)+\imu\Bigl[
   \tilde{\zeta} \Bigl(\tilde{\zeta}^2+
   {\textstyle \frac{1}{2}}(1+\bar k^2)\Bigr)
   D'(b'\tilde{\zeta})b'^{-2}\\[0mm]\nonumber
  &&\!\!{}-{\textstyle \frac{1}{8}} \Bigl(3\bar k'^4 + 8 \bar k^2\Bigr)
   \left(\bE'-E(\psi,\bar k')\right) \\[0mm]
  &&\!\!\!\left.\Bigl. {} + {\textstyle \frac{1}{16}}(1+\bar k^2)
   \Bigl(5\bar k'^4+8\bar k^2\Bigr) \nonumber
   \left(\bK'-F(\psi,\bar k')\right)\Bigr]\right\} 
  \quad,
\eea
and the value for $A^{(2)}(a',b')$ reads
\begin{eqnarray*}
  \!\!\!\!&&  A^{(2)}
  = A^{(1)}
  \{\tau_0^{(1)}\to\tau_0^{(2)},
  \tau_2^{(1)}\to\tau_2^{(2)}\} \\[0mm]
  \!\!\!\!&&{}+
  {\textstyle \frac{1}{40}}
  \tau_4^{(2)}b'^5\left[\frac{1+\bar{k}^2}{2}
  \Bigl(5\bar{k}'^4 + 8\bar{k}^2 \Bigr)\bK'
-\Bigl(3\bar{k}'^4+8\bar{k}^2\Bigr)\bE'\right]
  \,,
\end{eqnarray*}
Also, as for the second order in Sect.\ \ref{sect:esBp-1st-ord},
difference and sum of the equations (\ref{two_eqs:sec_order}) yield
\bea
  \!\!\!&& 0 = \tau_0^{(2)}+\thalf\tau_2^{(2)}b'^2(1+\bar{k}'^2)
   +{\textstyle\frac{1}{8}}\tau_4^{(2)}b'^4(3+2\bar{k}'^2+3\bar{k}'^4)
   \,,
   \nonumber \\ [2mm]
  \!\!\!&& 0 = 2V_f + {\textstyle\frac{2}{3}} \tau_2^{(2)}b'^4
   \left(2\bar k'^2\bK'-(1+\bar k'^2)\bE'\right)
   \nonumber \\[0mm]
  \!\!\!&&{}+ {\textstyle\frac{1}{5}} \tau_4^{(2)}b'^6
   \left(4\bar k'^2(1+\bar k'^2)\bK'
   -(3 + 2\bar k'^2+3\bar k'^4)\bE'\right)
   \,,
   \nonumber
\eea
where in the sum $A^{(2)}$ was inserted and $\tau_0^{(2)}$ was
taken from the difference.

\section{Potential configuration of Metallic Components}

\subsection{Closed form for the general potential}

\label{app:calc-metpot-gen}

The general solution for $\phi^{\rm el.}(x_2,y_2)$ for the capacitor represented in Fig.~\ref{inverted_system} (right), of $4\bK$ 
periodicity along the $x_2$ direction, reads $\phi^{\rm el.} = \tilde V_2 + (\tilde V_1-\tilde V_2)C/ 2\bK + 
\tilde V_0 y_2/\bK' +(\tilde V_1 -\tilde V_2)/(2\bK) f_\phi$, where
\bea
   f_\phi = 2\sum_{n=1}^\infty \frac{\sin(\alpha_nC)\sinh[\alpha_n(\bK'-y_2)]}
  {\alpha_n\sinh(\alpha_n\bK')}\cos(\alpha_nx_2) 
  \label{f_phi_def}
\eea 
with $\alpha_n=n\pi/2\bK$ and, separated in $f_\phi=f_\phi^{(1)}+ f_\phi^{(2)}$,
is given in (\ref{f_phi_1}) and (\ref{f_phi_2}).

The first term containing products of trigonometric functions and an
exponential can be written as
\bea
\hspace{-2mm}  f_\phi^{(1)}\hspace{-1mm} & = &\hspace{-1mm} -\frac{2\bK}{\pi} \RE\left\{\imu\sum_{n=1}^\infty
  \frac{\euz^{\imu\alpha_n(\zeta_2+C)}-
  \euz^{\imu\alpha_n(\zeta_2-C)}}{n}\right\} 
  \label{f_phi_1}
\\[2mm]
&=& -\, C+\frac{2\bK}{\pi}\RE\left\{\imu
  \ln  \frac{\sin\left(\frac{\pi}{4\bK}(\zeta_2+C)\right)}
  {\sin\left(\frac{\pi}{4\bK}(\zeta_2-C)\right)}\right\}  ,
\eea 
with $\zeta_2=x_2+\imu y_2$, 
and 
where, in general, depending on the arguments inside the logarithm,
a constant $4\bK\Delta n$ may be added.
The second term containing products of trigonometric functions and a
hyperbolic sine can be written as
\bea
  &&f_\phi^{(2)} = {}
   \label{f_phi_2}
  \\*
  &&\frac{8\bK}{\pi}\RE\left\{\imu
   \sum_{n=1}^\infty \frac{1}{n}
   \frac{\euz^{-2\alpha_n\bK'}}{1-\euz^{-2\alpha_n\bK'}}
   \sin(\alpha_n \zeta_2)\sin(\alpha_nC)\right\} \;.
  \nonumber
\eea
Since here just a product of two sine functions occurs, 
the following formula \cite{stegun} can be used:
\[
  4\sum\limits_{n=1}^\infty
  \frac{q^{2n}\sin 2n\alpha \sin 2n\beta}{n(1-q^{2n})}
  =\ln\frac{\vartheta_1(\alpha+\beta,q)}{\vartheta_1(\alpha-\beta,q)}-
  \ln\frac{\sin(\alpha+\beta)}{\sin(\alpha-\beta)} ,
\]
identifying $\alpha=\pi \zeta_2/4\bK$ and $\beta=\pi C/4\bK$.
Because $k$ and therefore $k'$, $\bK$, $\bK'$, and
$q=\exp\{-\pi\bK'/\bK\}$ are fixed, the variable $q$ related
to $f_\phi^{(2)}$ enters as $q\to \sqrt{q}$ into the sum
resulting in
\bea
  f_\phi^{(2)}&=&\frac{2\bK}{\pi}\RE\left\{\imu\ln\frac
  {\vartheta_1\left(\frac{\pi}{4\bK}(\zeta_2+C),
   \euz^{-\frac{\pi\bK'}{2\bK}}\right)}
   {\vartheta_1\left(\frac{\pi}{4\bK}(\zeta_2-C),
   \euz^{-\frac{\pi\bK'}{2\bK}}\right)}
   \right.\nonumber\\*
   &&\qquad\quad\left.
   \vphantom{\frac{\left(A^{\frac{G'}{H}}\frac{C}{D}\right)}
            {A^{\frac{G'}{H}}\frac{C}{D}}}
   {}-\imu\ln\frac
   {\sin\left(\frac{\pi}{4\bK}(\zeta_2+C)\right)}
   {\sin\left(\frac{\pi}{4\bK}(\zeta_2-C)\right)}
  \right\} \quad .
\eea

For $f_\phi=f_\phi^{(1)}+f_\phi^{(2)}$ representing
(\ref{f_phi_def}) the logarithmic sine terms cancel
and $\phi^{\rm el.}(\zeta_2,\bar \zeta_2)$ 
obtains the form (\ref{close_form}).

\vspace*{\bigskipamount}

{\it Limit to the Symmetric Case}---The limit in (\ref{close_form}) 
is performed by setting
by $C\to\bK$, simplifying the first
arguments of the $\vartheta_1$ functions.
Subsequently various formulae of Sect.\ 8.1 of Ref.~\onlinecite{gradshteyn}, are used.
The quantity $\sqrt{q}$ corresponds to the modulus
$k_\abbr{tr}=2\sqrt{k}/\allowbreak(1+k)$. After argument translation the
$\vartheta_1$ functions are rewritten in terms of elliptic functions
of modulus $k_\abbr{tr}$, which then is transformed to $k$.
The subsequent argument doubling and translation yields

\begin{eqnarray}
  \phi^{{\rm el.}(\rm s)}(x_2,y_2)
  &=& \tilde V_1+\left(V_\abbr{g}-\frac{\tilde V_1+\tilde V_2}{2}\right)
   \frac{y_2}{\bK'} \\* \nonumber
  &&{}+
   \frac{\tilde V_1-\tilde V_2}{2\pi}\RE\left\{\imu\ln
   \frac{1+\sn\left(\zeta_2,k\right)}{1-\sn\left(\zeta_2,k\right)}
   \right\} \;.
  \label{potential_sec_way}
\end{eqnarray}

This result for the symmetric case can be obtained in a simpler way,
where the limit $C\to\bK$ in (\ref{f_phi_def}) just leaves the odd $n$
with an alternating sign. There the summation, after a few manipulations
involving an intermediate derivation by $\zeta_2$, leads directly to
elliptic functions of modulus $k$.

The dynamical limit for $C\to\bK$, i.e., $\bar b\to -b, \bar a\to -a$,
results from $\tilde V_1=\tilde V_2$. The dependence of the edge
position $b$ on $(V_1 - V_\abbr{g})/(\tau a)$ is shown in Fig.\
\ref{schematic_illustration_2}.
 
\vspace{5mm}
\begin{figure}[htbp]
\begin{center}
\includegraphics[width=0.68\columnwidth]{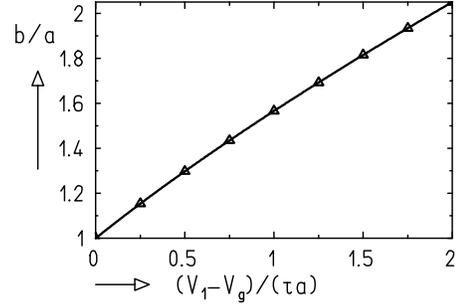}
\end{center}
\caption{Plot of edge position $b$ (see Fig.~\ref{schematic_illustration})
as a function of voltage $V_1 - V_\abbr{g}$ is shown. Lengths are measured in
units of half of the gate width, $a$, which here is taken to be
$a=50 \, l_{B}$.}
\label{schematic_illustration_2}
\end{figure}

\subsection{Equations Used for $E_{y_2}$}
\label{app:Ey2-eqs}

The expression (\ref{eq:Ey2-by-theta}) for $E^{\rm el.}_{y_2}(\zeta_2)$
is rephrased using the Jacobi zeta function $Z(u)$, Eqs.~(16.34.1,2) of Ref.~\onlinecite{stegun}, and
its relation Eq.~(17.4.28) of Ref.~\onlinecite{stegun} to elliptic integrals $E$ and elliptic functions.
Their modulus is $k_\abbr{tr}=2\sqrt{k}/(1+k)$, which is transformed to
$k$ for both function types leading to
\begin{widetext}
\be
  \label{eq:Ez2(z2)trf}
  E^{\rm el}_{y_2}(\zeta_2) =
  -\frac{\tilde V_0}{\bK'}  
+\frac{\tilde V_1-\tilde V_2}{2\pi}
  \RE\hspace{-1mm}\left[ 
     \vphantom{\frac{\left(\frac{A}{C}\right)}{\left(\frac{A}{C}\right)}}
  2E\left\{\am\left(\frac{\zeta_2+C}{2},k\right),k\right\}
  -(\zeta_2+C)\frac{\bE}{\bK} \right. 
  \left.{}+\frac{\cn\left(\frac{\zeta_2+C}{2},k\right)
  \dn\left(\frac{\zeta_2+C}{2},k\right)}
  {\sn\left(\frac{\zeta_2+C}{2},k\right)}-(C\to-C)\right] \;.
\ee
\end{widetext}
[Notice also that the arguments of the incomplete elliptic integral
of second kind $E(\cdot,\cdot)$ depend on the convention:
see Refs.~\onlinecite{gradshteyn} and \onlinecite{stegun}.]
Then using Eq.~(8.157) of Ref.~\onlinecite{gradshteyn} for half arguments and subsequently Eq.~(8.151 2.)\ of Ref.~\onlinecite{gradshteyn},
making the selection of the sign depending on the sign of the
elliptic functions noticing $0\le C \le 2\bK$.
With this and the argument shift for elliptic integrals, Eq.~(17.47) of Ref.~\onlinecite{stegun},
for the special arguments $\zeta_2=0,2\bK$, one gets the expressions
{\small
\bea
\label{E_y2_0}
  \!\!\!&&
  E^{\rm el.}_{y_2}(0) =
  -\frac{\tilde V_0}{\bK'} + \frac{\tilde V_1-\tilde V_2}{\pi}
  \Bigl[2E\left(\am\left(\thalf C\right)\right)-C\frac{\bE}{\bK}
  \Bigr. \nonumber\\*[0mm]
  \!\!\!&& 
  \Bigl. {} +
  k'\left[{\frac{1-\sn(C-\bK)}{1+\sn(C-\bK)}}\right]^{\frac{1}{2}}\Bigr] \;,
  \\
  \label{E_y2_2K}
  \!\!\!&&
  E^{\rm el.}_{y_2}(2\bK) =
  -\frac{\tilde V_0}{\bK'}
  + \frac{\tilde V_1-\tilde V_2}{\pi}
  \left[2E\left(\am\left(\thalf C\right)\right)-C\frac{\bE}{\bK}
  \right.
  \nonumber\\[0mm]
  \!\!\!&&
  \left.{}
  -2k\left[{\frac{\dn(C-\bK)-k'}{\dn(C-\bK)+k'}}\right]^{\frac{1}{2}}
  -k'\left[{\frac{1+\sn(C-\bK)}{1-\sn(C-\bK)}}\right]^{\frac{1}{2}}\right] \;,
  \qquad
\eea}%
where the second argument of $\am,E,\sn,\dn$ is $k$.

Starting from the expression for $s_1$
(Ref.~\onlinecite{s_12})
one obtains the shift $P$ of Eq.~(\ref{inv_1}).   
The modulus $k$ of the transformation is defined in (\ref{sn_transf}) as $I(b)/I(a)$;
its explicit form is given in Sect.\ \ref{sect:noi-fld-str}  
as $k=S_k^-/S_k^+$. 
From that the right-hand side of the equation $\sn(C-\bK,k)=P/I(b)$
can be written as
{\small
\be
 \label{eq:asy-k-const-gen_new}
  \frac{P}{I(b)}=\frac{\left[{(\bar a-b)(a-b)}\right]^{1/2}-\left[{(\bar a-\bar b)(a-\bar b)}\right]^{1/2}}
   {\left[{(\bar a-b)(a-b)}\right]^{1/2}+\left[{(\bar a-\bar b)(a-\bar b)}\right]^{1/2}} \;,
\ee}
which 
allows to express the following terms:
\be
  k'\left[{\frac{1\mp\sn(C-\bK)}{1\pm\sn(C-\bK)}}\right]^{\frac{1}{2}}=
  \frac{2\left[{( a_\mp- b_\mp)(a_\pm- b_\mp)}\right]^{\frac{1}{2}}}{S_k^+} \;, 
 \label{eq:k-sq-dn-terms-0}
\ee
\bea
 && k\left[{\frac{\dn(C-\bK)-k'}{\dn(C-\bK)+k'}}\right]^{\frac{1}{2}}={}
 \nonumber
 \\*
 &&\frac{b-\bar b-\left[{(\bar a-b)(a-b)}\right]^{\frac{1}{2}}-\left[{(a-\bar b)(\bar a-\bar b)}\right]^{\frac{1}{2}}}{S_k^+} \;.
 \qquad
 \label{eq:k-sq-dn-terms}
\eea
In (\ref{eq:k-sq-dn-terms-0}) we have used the short notation $a_\pm$, respectively for $a$, $\bar a$, and $b_\pm$, respectively 
for $b$, $\bar b$.


\begin{thebibliography}{50}
\bibitem{larkin_shikin} Larkin and Shikin (Ref.\ \onlinecite{larkin})
have solved a similar problem but for the situation of semi-infinite
metallic gates confining a 1D electron channel. For this we would have 
found a similar result to Eq.~(\ref{eq:position_equation}),
with the right-hand side substituted by $(k^2\bK'-\bE')/k$, which
agrees with theirs (after making the appropriate correspondence
$d^{\rm LS}\to b$, $V^{\rm LS}_1=0$, $V^{\rm LS}_\abbr{g} \to -V_\abbr{g}$).
\bibitem{D_F} The functions $D(\zeta)$ and ${\cal F}(\zeta)$ for
asymmetric dielectric stripes are
$ D(\zeta)=[(\zeta-\bar b)(\zeta-\bar a)(\zeta-a)(b-\zeta)]^{1/2}$
and 
$ {\cal F}(\zeta) =
\thalf(\zeta-\bar b)(\zeta-b)+\thalf(\zeta-\bar a)(\zeta-a)
-{\textstyle\frac{1}{8}}(b-a-\bar a+\bar b)^2$.
This ${\cal F}(\zeta)$ is valid only for uniformly charged stripes.
\bibitem{s_12} The condition $I(\bar a) - I(\bar b) = I( b) - I( a)$
is resolved at two points
$s_{1,2}= (\bar bb-\bar aa \allowbreak\pm \left[{(\bar b-\bar a)(\bar b-a)
\allowbreak(\bar a-b)(a-b)}\right]^{1/2} )/\allowbreak
  (\bar b-\bar a-a+b)$. We select for our calculations $s_1$ only.
\bibitem{length_l} The length $l$ that appears in (\ref{eq:lim-x-0}), (\ref{eq:lim-x-2K}) is
$ 
\textstyle   l = b-\bar b-2[{(a-\bar b)(\bar a-\bar b)}]^{\frac{1}{2}}
- S_k^+ \left(2E\left(\am\left(\thalf C\right)\right)-\frac{C\bE}{\bK} \right) .
$
\bibitem{factor} As long as is $k$ not small, in principle a factor
 $(1-k^2)/(1-k^2f^2)^{1/2}$ appears under the square-root.
\bibitem{identity}
In the text  to calculate correlation functions of a product of electron annihilation (creation) operators 
of the form $R\sim \eun^{{\rm i}\sqrt{4\pi}\phi(x)/\nu}$,  we use the identity
 \bea
   \left< \Pi_i\eun^{\hat{O}_i}\right> =
  \exp\left(\frac{1}{2}\left<2\sum_{i<j}\hat{O}_i\hat{O}_j +
  \sum_i\hat{O}^2_i\right>\right) \quad ,
 \eea
where $\hat{O}_i\sim {\rm i} \sqrt{4\pi}\phi(x)/\nu_i$
\end{thebibliography}
\end{document}